\documentclass[10pt]{wlscirep}
\usepackage[T1]{fontenc}
\usepackage{hyperref}
\usepackage{float} % for the [H] specifier
\usepackage{amsmath}
\usepackage{graphicx}
\usepackage{booktabs}
\title{First Positronium Lifetime Imaging using $^{52}$Mn and $^{55}$Co with a plastic-scintillator-based PET scanner}

\author[1,2,9]{Manish Das}
\author[1,2,*]{Sushil Sharma}
\author[1,2]{Ermias Yitayew Beyene}
\author[6]{Aleksander Bilewicz}
\author[3]{Jarosław Choiński}
\author[1,2]{Neha Chug}
\author[4]{Catalina Curceanu} 
\author[1,2]{Eryk Czerwi{\'n}ski}
\author[5]{Jakub Hajduga}
\author[1,2]{Sharareh Jalali}
\author[1,2]{Krzysztof Kacprzak}
\author[1,2]{Tevfik Kaplanoglu}
\author[1,2]{Łukasz Kapłon}
\author[1,2]{Kamila Kasperska}
\author[1,2]{Aleksander Khreptak} 
\author[1,2]{Grzegorz Korcyl}
\author[1,2]{Tomasz Kozik}
\author[1,2]{Karol Kubat}
\author[1,2]{Deepak~Kumar}
\author[1,2]{Sumit Kumar Kundu}
\author[1,2]{Anoop Kunimmal Venadan}
\author[7]{Edward~Lisowski}
\author[7]{Filip Lisowski}
\author[1,2]{Justyna Medrala-Sowa} 
\author[1,2]{Simbarashe Moyo}
\author[1,2]{Wiktor Mryka}
\author[1,2]{Szymon Nied{\'z}wiecki}
\author[1,2]{Anand Pandey}
\author[1,2]{Piyush Pandey}
\author[1,2]{Szymon Parzych}
\author[1,2,4,10]{Alessio Porcelli}
\author[5]{Bartłomiej Rachwał}
\author[1,2]{Martin Rädler}
\author[8]{Narendra Rathod}
\author[6]{Noman Razzaq}
\author[8]{Axel Rominger}
\author[8]{Kuangyu Shi}
\author[1,2]{Magdalena Skurzok}
\author[1,2]{Maciej Słotwiński}
\author[3]{Anna~Stolarz}
\author[5]{Tomasz Szumlak}
\author[1,2]{Pooja Tanty}
\author[1,2]{Keyvan~Tayefi~Ardebili}
\author[1,2]{Satyam Tiwari}
\author[1,2]{Kavya~Valsan~Eliyan} 
\author[6]{Rafał~Walczak}
\author[1,2,S]{Ewa Ł. Stepień}
\author[1,2,S,**]{Paweł Moskal}

\affil[1]{Faculty of Physics, Astronomy and Applied Computer Science, Jagiellonian University, S.~{\L}ojasiewicza 11, 30-348 Krakow, Poland}
\affil[2]{Center for Theranostics, Jagiellonian University, Kopernika 40, 31-501 Krakow, Poland}
\affil[3]{Heavy Ion Laboratory, University of Warsaw, 00-927 Warsaw, Poland}
\affil[4]{Laboratori Nazionali di Frascati, INFN, Via E. Fermi 40, 00044 Frascati, Italy}
\affil[5]{AGH University of Krakow, 30-059 Kraków, Poland.}
\affil[6]{Center of Nuclear Chemistry and Radiochemistry, Institute of Nuclear Chemistry and Technology, 03-195 Warsaw, Poland}
\affil[7]{Cracow University of Technology, Faculty of Mechanical Engineering, al. Jana Pawła II 37, 31-864 Kraków, Poland}
\affil[8]{Department of Nuclear Medicine, Inselspital, Bern University Hospital, University of Bern, 3010, Bern, Switzerland}
\affil[9]{Doctoral School of Exact and Natural Sciences, Jagiellonian University, S.~{\L}ojasiewicza 11, 30-348 Krakow, Poland}
\affil[10]{Center of Astronomical Research, Technology, Education, and Outreach, University of Antofagasta, Avda. U. de Antofagasta 02800, 1240000 Antofagasta, Chile}

\affil[*]{corresponding author, e-mail: \url{sushil.sharma@uj.edu.pl}}
\affil[**]{corresponding author, e-mail: \url{p.moskal@uj.edu.pl}}
\affil[S]{Senior authors}

\keywords{PLI, Positronium, J-PET}

\begin{abstract}
This study demonstrates the applicability of $^{52}$Mn and $^{55}$Co radionuclides for positronium imaging. Positronium Lifetime Imaging (PLI) extends positron emission tomography by using the lifetime of positronium atoms as a probe of tissue molecular architecture. However, its practical use requires $\beta^{+}$ emitters that also provide an additional prompt $\gamma$ ray to mark the positron creation time. In this work, we report the first PLI measurements performed with $^{52}$Mn and $^{55}$Co using the modular J-PET. Four samples were studied in each experiment: two Certified Reference Materials (polycarbonate and fused silica) and two human tissues (cardiac myxoma and adipose). The selection of PLI events was based on the registration of  two 511~keV annihilation photons and one prompt gamma in triple coincidence. From the resulting lifetime spectra we extracted the mean ortho-positronium lifetime $\tau_{\text{oPs}}$ and the mean positron lifetime $\Delta T_{\text{mean}}$ for each sample. The measured values of $\tau_{\text{oPs}}$ in polycarbonate using both isotopes matches well with the certified reference values. Furthermore, $^{55}$Co reproduced identical results for fused-silica measurements at their respective uncertainty levels. In contrast, measurements with $^{52}$Mn in fused silica show a minor deviation, which could be caused by the Parafilm spacer. In myxoma and adipose tissue, the reduced $\tau_{\text{oPs}}$ values are mainly linked to the long storage history of the samples rather than to the choice of isotope. Comparing peak-to-background ratios and spectral purity, $^{55}$Co provides cleaner PLI data under the same experimental conditions. Although $^{52}$Mn offers a longer half-life and a multi gamma cascade enhancing $\beta^{+}$ + $\gamma$ coincidences, but at the expense of higher background. In this study, we demonstrate that the applied selection criteria on the data measured with the modular J-PET can be used for PLI studies even with radionuclides with complex decay patterns. 
\end{abstract}
\begin{document}
\thispagestyle{empty}
\flushbottom
\maketitle
\section*{Introduction}

Positron Emission Tomography (PET) is a well established medical modality for tracing and mapping metabolic activity \textit{in-vivo} within the human body. In a PET scan, a positron-emitting radiotracer is administered intravenously. The emitted positron annihilates with surrounding electrons, which produces a pair of 511-keV photons emitting in nearly opposite directions. Registering both photons in a coincident window defines the Line Of Response (LOR), and intersection of LORs acquired from the full scan are used to reconstruct the spatial map of annihilations sites. This technique enables the acquisition of precise non-invasive images of biological processes on a molecular scale ~\cite{Phelps661,Alavi:2021}. The conventional PET method quantifies tracer uptake values, but does not directly provide information about the tissue microenvironment. Positronium imaging, an emerging extension of PET, complements the use of conventional PET~\cite{IEEEpositronium-imaging, Moskal2025IEEE, EJNMMI2020, PMB2019, doi:10.1126/sciadv.abh4394}. During PET, in nearly 40\% of cases, a metastable bound state known as the positronium atom (Ps) is formed within the human body. This previously underexplored phenomenon can be measured and characterized \textit{in-vivo}~\cite{doi:10.1126/sciadv.adp2840}, enabling Positronium Lifetime Imaging (PLI) as a microenvironment sensitive extension of PET.
Ps is formed when a positron from a radiotracer binds with an electron from a nearby molecule. It can be formed in either of the two spin configurations: para-positronium (pPs), a short-lived state with a mean lifetime of 125~ps and ortho-positronium (oPs), a significantly longer lived state with mean lifetime of about 142~ns in vacuum~\cite{RMP2023}.The oPs experiences a substantial decrease in its mean lifetime when it interacts with biological molecules in tissue environment~\cite{Ahn-Gidley-colagen-2021, doi:10.1126/sciadv.abh4394, Jean2006, Chen2012, Jasinska2017, Zgardzinska2020, Karimi2023, EJNMMI2023, Moyo2022Feasibility}. Where positron can picks-off an electron from surrounding molecules, annihilating mainly into two photons~\cite{RMP2023,jean:2003principles}. Therefore, pick-off rate depends on the electron density and size of the cavity making the oPs mean lifetime a sensitive probe of nanoscale free-volume structure. Another important factor is the oxygen concentration in the tissues which can further reduce the mean lifetime~\cite{Shibuya2020, Stepanov2020}. Oxygen molecules are paramagnetic in nature and during collision with oPs, they induce conversion from oPs to pPs. The third process is the oxidation which refers to the chemical quenching of oPs via electron transfer to different reactive species (e.g., free radicals, certain biomolecules, oxidants), forming the transient chemical complexes with much shorter lifetimes ~\cite{RMP2023, Gidley2006}. 
Consequently, measuring the mean lifetime of oPs in biological tissues allows for a direct and quantitative measure to study the contrast of tissue microenvironment and hypoxia~\cite{Shibuya2020, Stepanov2020, BAMShypoxia2021}. This makes Ps lifetime imaging (PLI) a promising biomarker for tissue structure analysis for disease characterization.

The technique has become widely recognized and has stimulated a significant interest in the implementation of PLI in clinical PET scans. This has led to new development in both clinical and preclinical scanners 
worldwide~\cite{doi:10.1126/sciadv.adp2840, doi:10.1126/sciadv.abh4394, Huang2025Fast, Steinberger2024, Mercolli2024.10.19.24315509, Prenosil-quadra-2022, Dai-PennPET-NEMA, Huang_mic, Karp136, Huangjnumed.125.270130, Samanta2023Feasibility, Takyu_2024, TAKYU2024169514, Mercolli2025}. 
However, the selection of a suitable radionuclide remains the main challenge limiting PLI's clinical applications~\cite{ManishBAMS, Moskal2025IEEE}. 
The ideal radioisotope needs to produce an additional prompt gamma emitted immediately after positron emission.
In PLI, the registration time of the prompt gamma is used as a precise timestamp of positron creation, which serves as the start time for the positronium lifetime measurement. Subsequently, detection of the annihilation photons from direct annihilation of the positron with surrounding electrons or through the formation of the Ps atom allows measurement of the positron lifetime. For practical reasons, the radioisotope for clinical PLI should meet additional criteria, which include a high branching ratio for $\beta^++\gamma$ emission, a half-life compatible with PET scans, and manufacturing via standard protocols~\cite{Sitarz2020, ManishBAMS, Moskal2025IEEE}. 
%--------

The proof of concept for PLI in biological tissue was demonstrated by the J-PET collaboration, which performed the first \textit{ex-vivo} PLI experiment using the J-PET tomograph~\cite{doi:10.1126/sciadv.abh4394}. In this study, $^{22}$Na was used as the positron 
source, well suited for \textit{ex-vivo} PLI emitting a 1275~keV prompt gamma in 99.94\% of $\beta^+$ decays. For this study, special phantoms were prepared with different biological tissues (cardiac myxoma, adipose) excised from two different patients. In each phantom, a $^{22}$Na source was sandwiched between tissues samples. All phantoms were then placed in the J-PET geometry. A clear difference in oPs mean lifetime was observed between cardiac myxoma (1.9~ns) and adipose tissues (2.6~ns)~\cite{doi:10.1126/sciadv.abh4394}. In contrast to earlier studies using $^{22}$Na for conventional positron annihilation lifetime spectroscopy yielding only bulk-mean lifetimes for small samples, J-PET collaboration provided for the first time spatially resolved o-Ps lifetime imaging, directly demonstrating the potential of PLI to distinguish tissue types based on their molecular architecture. However, the long half-life of $^{22}$Na (2.60 years) and its accumulation in bones~\cite{Triffitt1970boneuptake} preclude its uses for \textit{in-vivo} applications. Nevertheless, it remains primarily a source of choice for in-vitro and \textit{ex-vivo} PLI studies~~\cite{Ahn-Gidley-colagen-2021, doi:10.1126/sciadv.abh4394, Jean2006, Chen2012, Jasinska2017, Zgardzinska2020, Karimi2023, EJNMMI2023, Moyo2022Feasibility, KubatBAMS2024}, as well as for various fundamental physics experiments~\cite{Moskal:2021kxe, Moskal2025Nonmaximal, Moskal2024Discrete}. 
Recently, J-PET collaboration performed \textit{in-vivo} PLI of the human brain in a patient injected with $^{68}$Ga-PSMA and $^{68}$Ga-DOTA-Substance-P by means of the Modular prototype of the J-PET scanner~\cite{doi:10.1126/sciadv.adp2840}. Although the study was with low statistics, it showed the feasibility of applying PLI in clinical applications by successfully demonstrating imaging of the mean lifetime of the oPs~\cite{doi:10.1126/sciadv.adp2840}. One of the principal reason for low statistics is the low branching ratio (1.34\%) of prompt gamma and positron emission in $^{68}$Ga. 
\begin{figure}[ht]
\centering
\includegraphics[width=\linewidth]{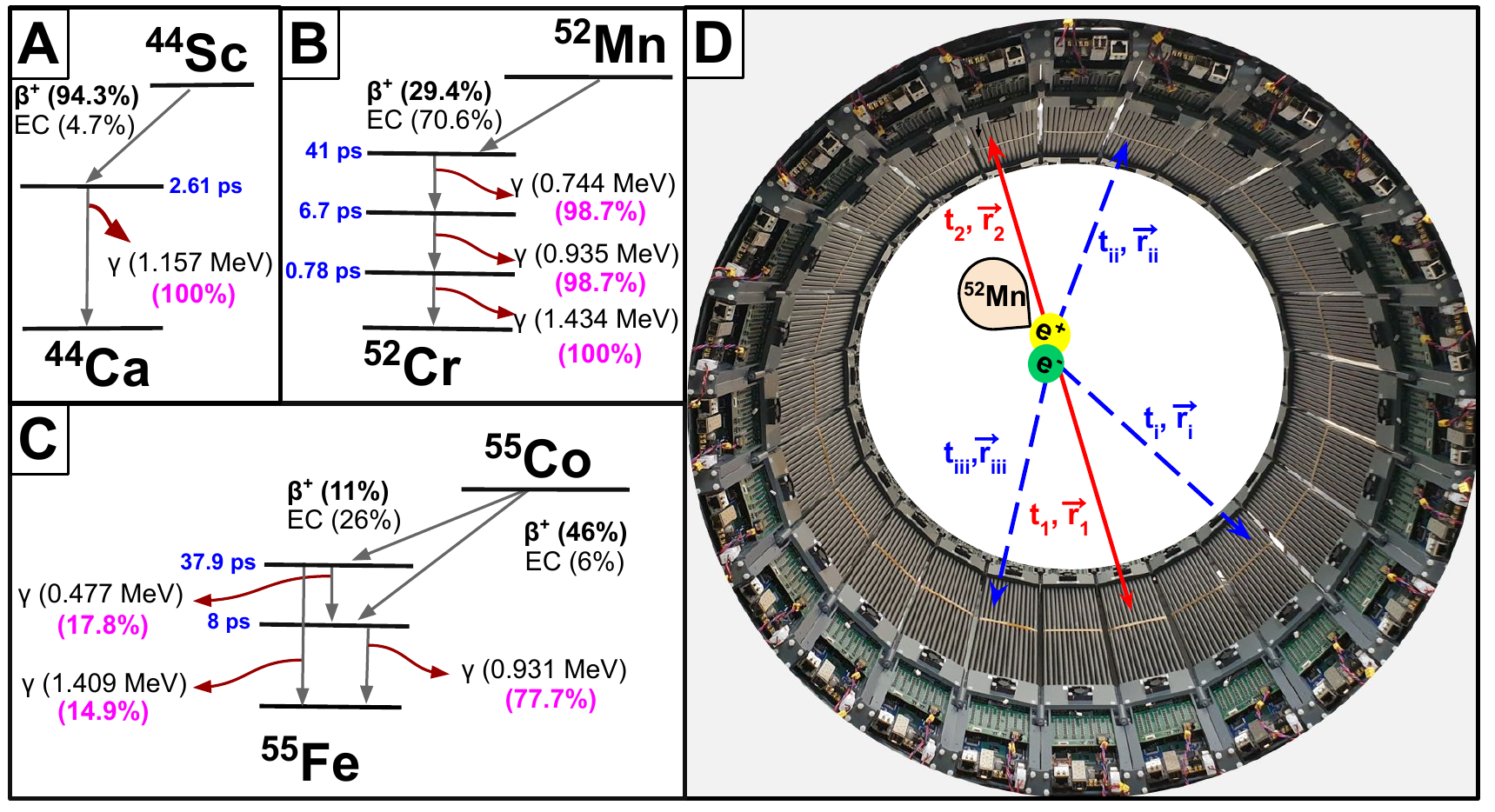}
\caption{\textbf{(A-C)} Decay scheme of $^{44}$Sc (A, $T_{1/2}=4.04$~h), $^{52}$Mn (B, $T_{1/2}=5.59$~d), and $^{55}$Co (C, $T_{1/2}=17.53$~h), where $\beta^{+}$ represents the positron emission yield, EC denotes electron capture contributions, $\gamma$ indicates the prompt gamma ray with its energy shown in parentheses, and the number in pink parentheses shows the fraction of $\beta^{+}+\gamma$ decay relative to all $\beta^{+}$ decays for a given isotope. The delay time, shown in blue text for clarity, represents the average interval between positron emission and prompt gamma emission. \textbf{(D)} Event definition for $^{52}$Mn in the modular J-PET scanner. Two annihilation photons ($t_1$, $\vec{r}_1$) and ($t_2$, $\vec{r}_2$), with a possible cascade of three prompt photons ($t_{i}$, $\vec{r}_{i}$), ($t_{ii}$, $\vec{r}_{ii}$), ($t_{iii}$, $\vec{r}_{iii}$). } 
\label{fig:decay_scheme}
\end{figure}
The successful proof-of-principle studies using $^{68}$Ga along with validation tests on PET/CT systems with long axial field-of-view~\cite{Mercolli2025} have led to efforts to explore additional clinically available $\beta^+$ emitters that produce prompt photon emissions for PLI. 

Two such isotopes are cycltron produced $^{124}$I, frequently used for PET/CT thyroid staging and dosimetery purposes, and generator produced $^{82}$Rb to perform PET perfusion imaging for the evaluation of blood flow of coronary artery disease. Recently, positronium lifetime imaging with $^{124}$I was demonstrated using Biograph Vision Quadra at Bern. The studies were performed for phantoms as well as  \textit{in-vivo} thyroid cancer patient imaging~\cite{Mercolli2025,Mercolli2025IODINE}. Similarly, $^{82}$Rb was used to conduct phantom experiments by PennPET Explorer at Philadelphia~\cite{Huangjnumed.125.270130}, while Biograph Vision Quardra at Bern was used to perform \textit{in-vivo} imaging of patients~\cite{Mercolli2024.10.19.24315509}. Although, both isotopes are suitable for initial PLI research but for effective application there are the notable challenges. The short half-life of $^{82}$Rb (1.26 minutes) and its low prompt gamma emission rate (13$\%$) in $\beta^+$ decays constrains the achievable statistics. On the other hand, $^{124}$I decay scheme produces a prompt gamma emission through 12\% of its $\beta^+$ decays and faces the same limitation. 
In recent years, these experiments with $^{68}$Ga, $^{124}$I and $^{82}$Rb revealed the constraints that PLI imposes on the underlying properties of radionuclides. 
Along the ongoing efforts to find optimal radioisotope, $^{44}$Sc stands out as one of the most promising candidate for PLI~\cite{pet-clinics}, due to the clinically relevant half-life of 4.04~hours~\cite{Duran2022Half}, an ultrashort de-excitation delay of 2.61 picoseconds, and a high yield of 94.3\% of decays resulting in positron emission followed by a high-energy prompt gamma of 1157 keV with a 100\% yield as shown in the Fig.\ref{fig:decay_scheme}(A). Recent phantom studies conducted with the PennPET Explorer in Philadelphia~\cite{Huangjnumed.125.270130}, the J-PET scanner in Cracow, Poland~\cite{ManishIEEE}, and the Biograph Vision Quadra in Bern~\cite{MercolliSc} have further demonstrated its potential as a leading candidate for future clinical applications of PLI. Although $^{44}$Sc fulfills many requirements for PLI, its relatively short half-life does not allow for use in radiopharmaceuticals with slow pharmacokinetics, such as monoclonal antibodies. The centers around the world have access to different production routes (e.g., generator based vs cyclotron based, low vs high current beams), which can affect the frequent availability of isotopes. Therefore, it is necessary to search for alternative $\beta^+$+$\gamma$ emitters. Several isotopes ($^{52}$Mn, $^{55}$Co, $^{60}$Cu, $^{72}$As, $^{14}$O, and $^{10}$C) have been proposed for their physical properties relevant to PLI~\cite{ManishBAMS, Moskal2025IEEE}.  

In this study, we present the first-ever results using two novel radioisotopes, Manganese-52 ($^{52}$Mn) and Cobalt-55 ($^{55}$Co), and their potential application in PLI. $^{52}$Mn, with a half-life of 5.59 days, emits positrons in 29.4\% of its decays, followed by a cascade of three prompt photons 744 keV, 935 keV , and 1434 keV as shown in Fig.~\ref{fig:decay_scheme}(B). This unique triple-photon cascade can significantly enhance the PLI sensitivity. Similarly, $^{55}$Co, with a half-life of 17.53 hours, produces positrons in 76\% of its decays, leading to excited $^{55}$Fe states that emit prompt gamma rays at 477 keV, 931 keV, and 1409 keV as shown in Fig.~\ref{fig:decay_scheme}(C), making it a good candidate for PLI.
Using these isotopes, we report the first experimental demonstration of  PLI with $^{52}$Mn and $^{55}$Co carried out on the Certified Reference Materials and the human tissues with the modular J-PET scanner~\cite{Moskal2024vision,FaranakBAMS2024,doi:10.1126/sciadv.adp2840}. The main objective of this study was to evaluate the capabilities and usability of the modular J-PET scanner for medical imaging using various $\beta^++\gamma$ isotopes using a data analysis protocol with voxel-based positronium imaging.
\section*{Methods}
\subsection*{Isotope preparation}\label{subsec0}
For this study, two separate experiments were conducted with two novel positron-emitting radionuclides, $^{52}$Mn and $^{55}$Co. The $^{52}$Mn isotope was produced under the PRISMAP - The European medical isotope programme at the Hevesy Laboratory in the Department of Health Technology at the Technical University of Denmark (DTU) via the $^{52}$Cr(p,n)$^{52}$Mn production route~\cite{Wooten2015Mnproduction} by irradiating a natural chromium (Cr) target for 8 hours at 40 µA current with a proton beam energy of 16.5 MeV on a GE PETtrace cyclotron. After irradiation, the activity was obtained in the Mn$^{2+}$ cation form in a 0.1 M HCl solution and transported to the Jagiellonian University in Krakow.  
$^{55}$Co was produced at the Heavy Ion Laboratory of the University of Warsaw via the $^{54}$Fe(d,n)$^{55}$Co reaction~\cite{Co_production, Cobalt_production}. A metal ($^{54}$Fe) target was irradiated for 2 hours with an 8.4 MeV deuteron beam at a current of 14.1 $\mu$A. After irradiation, the target was dissolved in 6 M HCl, subsequently filtered and $^{55}$Co was chemically separated at the Institute of Nuclear Chemistry and Technology in Warsaw. The solution with $^{55}$Co was then transported to the Krakow lab for experiments.
\subsection*{Sample preparation}\label{subsec1}
In each experiment, four different samples were examined: human adipose tissue, human cardiac myxoma tissue, and two Certified Reference Materials (CRMs) of fused silica and polycarbonate with known oPs lifetime. The CRMs were purchased from the National Institute of Advanced Industrial Science and Technology (AIST) in Tokyo, Japan. The adipose and myxoma tissues were selected as representative of human fatty tissue and tumor tissue, whereas the fused silica and polycarbonate CRMs served as reference materials with well-characterized oPs lifetime for calibration and for validating the stability of the measurement. 
%. 
\begin{figure}[H]
\centering
\includegraphics*[width=\linewidth]{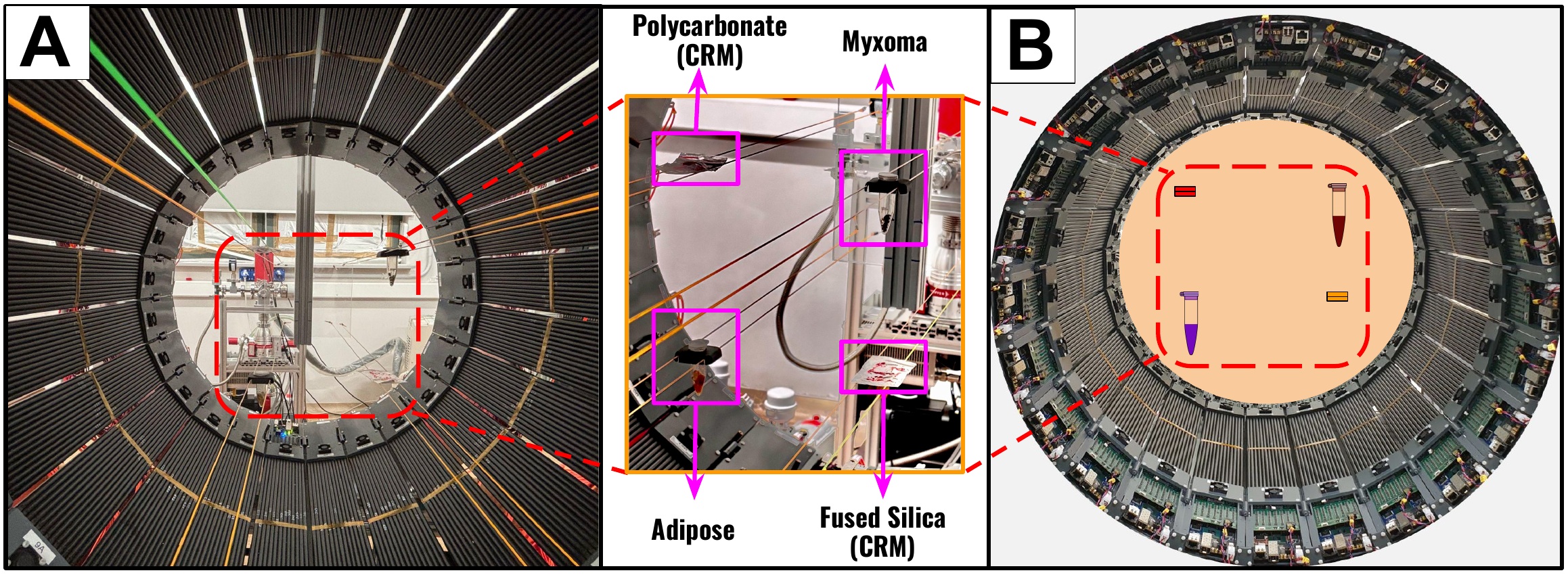}
\caption{ \textbf{(A)} Placement of the 4 samples inside the Modular J-PET scanner \textbf{(B)} Schematic representation of the 4 samples in the modular J-PET detector }
\label{fig:method}
\end{figure}
Cardiac myxoma and adipose tissue samples were obtained from a patient at John Paul II Hospital in Kraków, Poland, under bioethical consent number 1072.6120.123.2017. These tissues were previously used to demonstrate the first \textit{ex-vivo} positronium imaging with the J-PET scanner, as reported in ~\cite{doi:10.1126/sciadv.abh4394} and were re-used in the present work for positronium lifetime measurements with $^{52}$Mn and $^{55}$Co. After the previous experiment, the tissues were preserved in a formaldehyde solution, and stored till used for experiments. Before the current measurements, the tissue samples were removed from formaldehyde, stained with pure Lugol's solution for soft-tissue contrast for 24~hours at 40~$^\circ$C and drained~\cite{kubat2026exvivopositroniumimagingtissues}. At the time of measurements involving $^{52}$Mn, the tissue had been stored in the refrigerator for approximately three months, while at the time of measurements using $^{55}$Co the storage time was about five months. For the experiments, the adipose tissue and cardiac myxoma samples were placed in 5 mL Eppendorf tubes.  A 0.2 mL thin-wAll tube containing a solution of $^{52}$Mn (or $^{55}$Co) was inserted between the samples to prevent contamination. For preparing the CRM samples, the fused silica and polycarbonate plates supplied by AIST were used as received, only the source holder and mounting were prepared in-house for the experiment. In each of the CRMs, ten layers of Parafilm with a total thickness of approximately 1.2 mm were stacked. A small circular well was created in the center of the Parafilm layers to hold the radioactive source. The liquid source was placed in the well and sandwiched between two CRM plates. The dimensions of each CRM fused silica plate were $15~~\text{mm} \times 15~~\text{mm} \times 1.5~\text{mm}$. In case of polycarbonate each plate was of size $15~\text{mm} \times 15~\text{mm} \times 2~\text{mm}$. To ensure the containment of the source in the prepared well and prevent leakage, the entire assembly was securely wrapped with an additional single layer of Parafilm.

For the measurements with $^{52}$Mn, four different activities of 1.278 MBq, 1.298 MBq, 1.346 MBq, and 1.354 MBq, measured at the start of the experiment, were added to the samples prepared for cardiac myxoma, human adipose tissue, polycarbonate, and fused silica, respectively. The data were then collected for 19 hours 44 minutes with all four samples placed in the scanner (see Fig.~\ref{fig:method}A). In case of $^{55}$Co, activities of 1.032 MBq, 1.020 MBq, 1.092 MBq, and 1.071 MBq, measured at the start of the experiment, were used for cardiac myxoma, human adipose tissue, polycarbonate, and fused silica, respectively (see Fig.~\ref{fig:method}B). These measurements lasted 15 hours and 5 minutes. Apart from the choice of radionuclide and activity, the set-up was the same in both measurements, see Figs.~\ref{fig:method}(A-B).

\subsection*{Ethics Statement }\label{subsec7}

The study was conducted in accordance with the principles of the Declaration of Helsinki. It was approved by the Bioethical Committee of the Jagiellonian University (approval number: 1072.6120.123.2017). Informed consent was obtained from the patients.
\subsection*{Modular J-PET scanner and data processing }\label{subsec3}
In PLI, events are selected by requiring a triple coincidence of two nearly back-to-back 511~keV annihilation photons ($\gamma_a$) and an additional prompt gamma ($\gamma_p$), which defines the start time for estimating the positronium formation time. The corresponding selection criterion has been developed and validated in our previous studies~\cite{doi:10.1126/sciadv.abh4394, doi:10.1126/sciadv.adp2840, ManishIEEE} and is applied consistently in the present analysis. 
\begin{figure}[t]
\centering
\includegraphics[width=\linewidth]{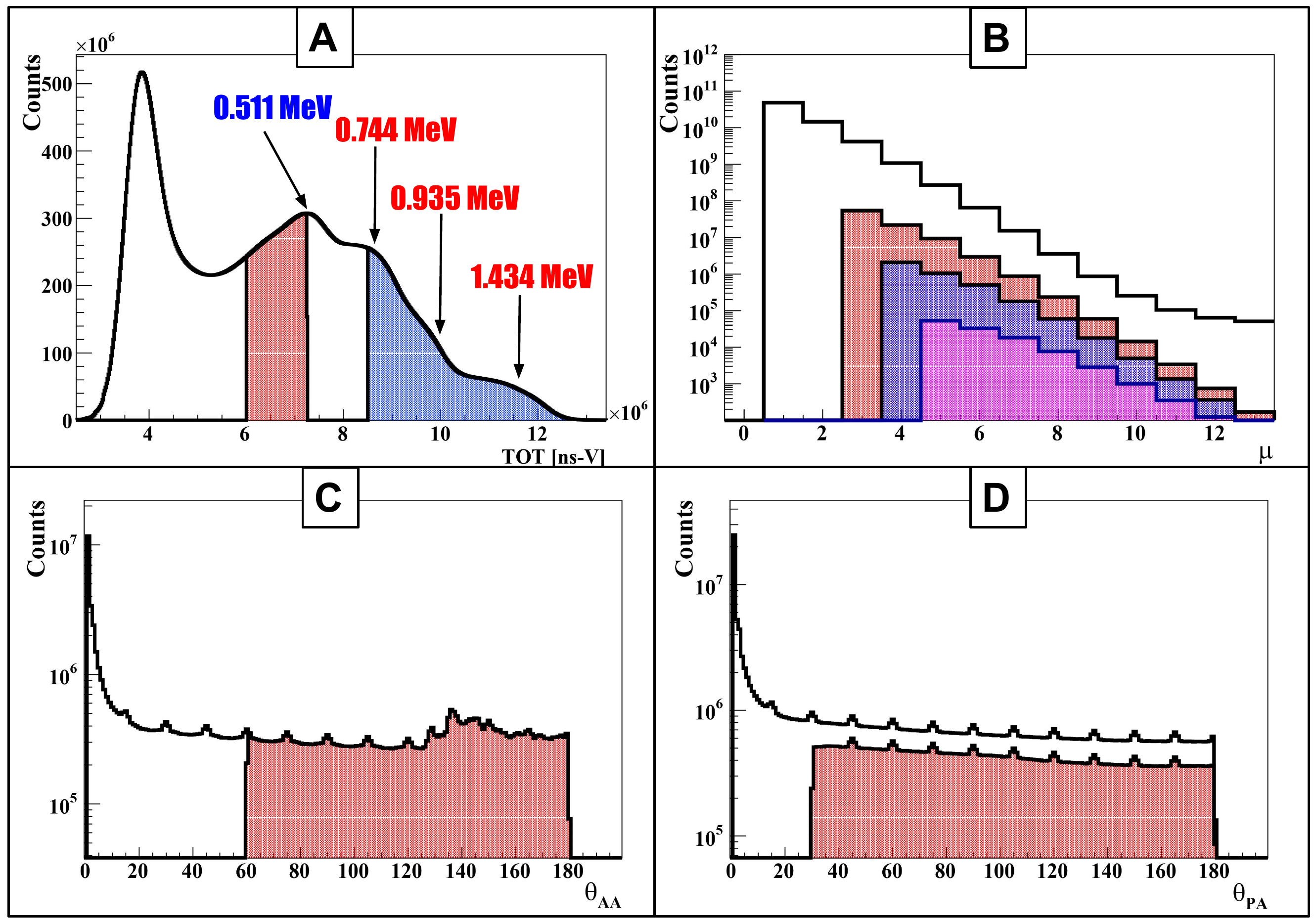}
\caption{Event Selection for $^{52}$Mn: {\bf{(A)}} Distribution of time-over-threshold (TOT$_{Hit}$) for photon identification, with annihilation photons (red) and prompt gammas (blue) marked by distinct ranges. {\bf{(B)}} The hit multiplicity ($\mu$) distribution for events is represented by histograms: the red-shaded histogram highlights events with exactly two annihilation photons and one prompt gamma, the blue-shaded histogram indicates events with exactly two annihilation photons and two prompt gammas, and the pink-shaded histogram denotes events with exactly two annihilation photons and three prompt gammas. {\bf{(C)}} Distribution of the relative angle ($\theta_{AA}$) between annihilation photon vectors $\vec{r}_1$ and $\vec{r}_2$ (per Fig.~\ref{fig:decay_scheme}D), with $\theta_{AA} \geq 60^\circ$ (red) as the selection criterion. {\bf{(D)}} Distribution of the relative angle ($\theta_{PA}$) between prompt gamma vector $\vec{r}_i$ and annihilation photon vectors $\vec{r}_1$, $\vec{r}_2$ (per Fig.~\ref{fig:decay_scheme}D), with $\theta_{PA} \geq 30^\circ$ (red) as the restriction. }
\label{fig:selection_mn}
\end{figure}
The modular J-PET scanner is an assembly of 24 detection modules arranged in a cylindrical configuration. Each module consists of 13 closely packed plastic scintillators of length 50 cm and a cross-section of 2.4~cm$\times$0.6~cm at the edges~\cite{doi:10.1126/sciadv.adp2840, FaranakBAMS2024, das2024}. Scintillators are read out from both ends by a 1$\times$4 matrix of Silicon Photomultipliers (SiPMs). J-PET DAQ operates in a continuous, triggerless data acquisition mode, which allows simultaneous registration of multiple photons from the same decay~\cite{korcyl_ieee, KorcylBAMS2014}. In plastic scintillators, photon interactions are mainly dominated by the Compton scattering mechanism, which leads to a continuous energy deposition determined by the scattering angles~\cite{NIM2014}. The signals recorded in the SiPM matrices are read out by the dedicated front-end boards and processed by an FPGA-based DAQ architecture built on the TRB3 platform. For each measured pulse, the leading and trailing edges are time stamped on a set of programmable thresholds. These times are further used to reconstruct the hit-time, hit-position, and the time-over-threshold (TOT) of the hit~\cite{korcyl_ieee,Niedzwiecki:2017nka, BAMSPalka2014}. In our analysis, we use the measured TOT as an estimator of the deposited energy~\cite{Sharma:2020}. For every registered interaction (hit), the TOT value is obtained as the average of the TOT measurements from the four SiPM matrices coupled to both ends of the strip.  

\begin{figure}[ht]
\centering
\includegraphics*[width=\linewidth]{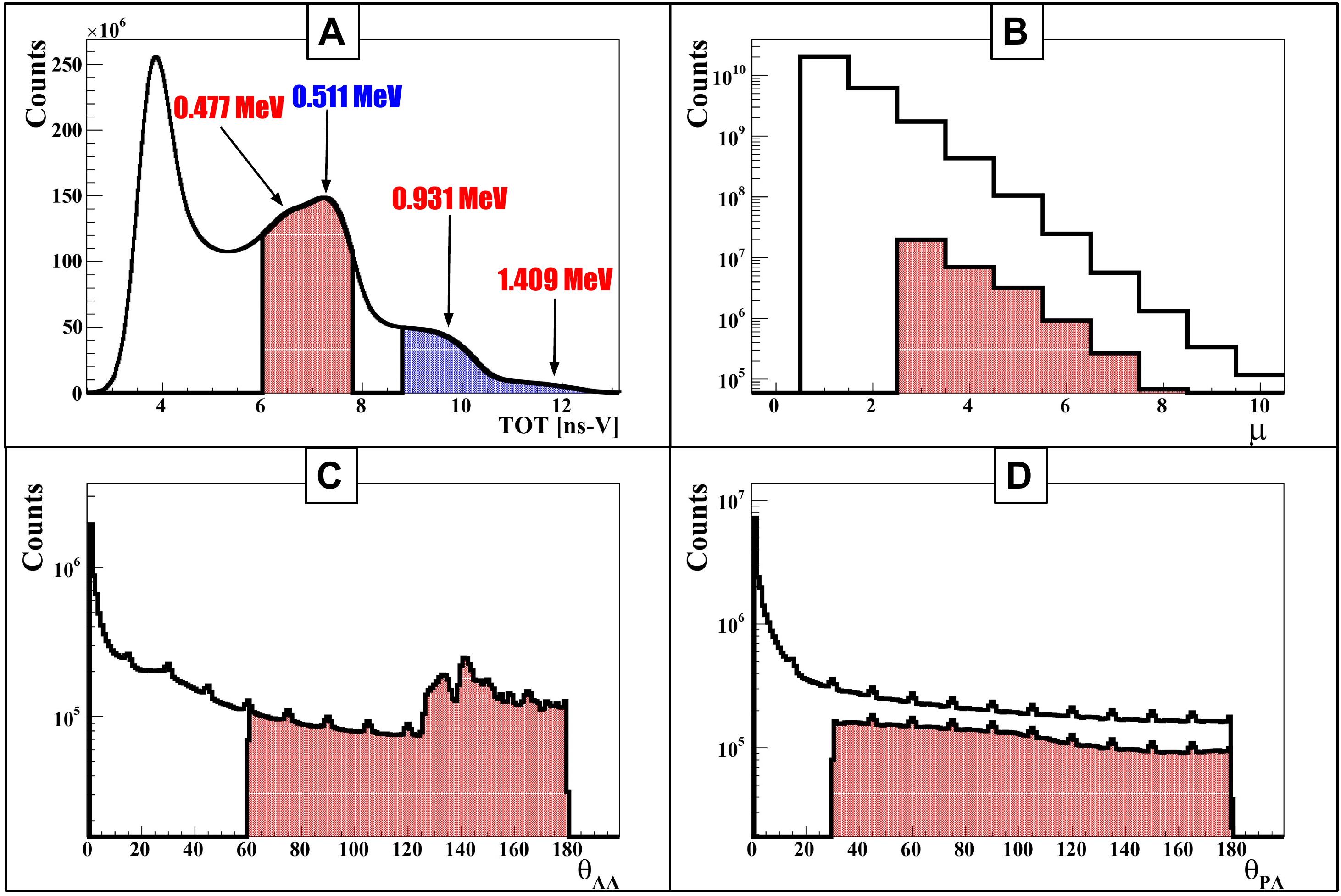}
\caption{Event Selection for $^{55}$Co: {\bf{(A)}} Distribution of time-over-threshold (TOT$_{hit}$) for photon identification, with annihilation photons (red) and prompt gammas (blue) marked by distinct ranges. {\bf{(B)}} Hit multiplicity ($\mu$) distribution for events, with the red-shaded histogram highlighting selected events containing exactly two annihilation photons and one prompt gamma. {\bf{(C)}} Distribution of the relative angle ($\theta_{AA}$) between annihilation photon vectors $\vec{r}_1$ and $\vec{r}_2$ (per Fig.~\ref{fig:decay_scheme}D), with $\theta_{AA} \geq 60^\circ$ (red) as the selection criterion. {\bf{(D)}} Distribution of the relative angle ($\theta_{PA}$) between prompt gamma vector $\vec{r}_i$ and annihilation photon vectors $\vec{r}_1$, $\vec{r}_2$ (per Fig.~\ref{fig:decay_scheme}D), with $\theta_{PA} \geq 30^\circ$ (red) as the restriction. }
\label{fig:selection_co}
\end{figure}

The measured TOT spectra for $^{52}$Mn and $^{55}$Co are shown in Fig.~\ref{fig:selection_mn}(A) and Fig.~\ref{fig:selection_co}(A), respectively. For $^{52}$Mn, the enhancements associated with the Compton edges of photons with energies 511, 744, 935, and 1434~keV are observed at values of 7.5, 8.5, 10 and 12~ns$\cdot$V, respectively (Fig.~\ref{fig:selection_mn}(A)). In case of $^{55}$Co, the Compton edges for photons with energies at 477, 511, 931, and 1409 keV appear at TOT values corresponding to 6.5, 7.5, 10 and 12~ns$\cdot$V, respectively (Fig.~\ref{fig:selection_co}(A)). These enhancements correspond to the maximum energy (Compton edge) that can be transferred to an electron in a single Compton scattering process.
For selecting the annihilation candidates (511 keV photons, red-shaded area), TOT values in the ranges 6 to 7.25 ns$\cdot$V for $^{52}$Mn and 6 to 7.8 ns$\cdot$V for $^{55}$Co were selected. For identifying the prompt photons (blue-shaded area), TOT value ranges of 8.5 to 13.5 ns$\cdot$V for $^{52}$Mn and 8.8 to 13.5 ns$\cdot$V for $^{55}$Co were used. Finally, event formation was based on the number of photon interactions (hits) occurring within a 20 ns coincidence window for both annihilation photons ($\gamma_a$) and prompt gamma photons ($\gamma_p$).
The standard PLI is based on the triple coincidence approach in which an event is composed of two annihilations and only one prompt photon. The decay process of $^{52}$Mn and $^{55}$Co produces several prompt photons that emerge simultaneously in time (see Fig.~\ref{fig:decay_scheme}B-C). In principle, each of these prompt photons can be used as a start signal for estimating the lifetime spectra. In this study, we explored whether prompt multiplicity can provide an additional practical advantage for PLI or not. For this, events with different hit multiplicities were studied. Here, the hit multiplicity is defined as the total number of photon interactions in an event. The event selection criterion used in this study is presented in Fig.~\ref{fig:selection_mn}(B) for $^{52}$Mn and Fig.~\ref{fig:selection_co}(B) for $^{55}$Co. In studies with $^{52}$Mn, we analyzed three different classes of events with hit multiplicities of three, four, and five, which correspond to including one, two, or three prompt photons appearing in the cascade, respectively. These events contain two 511~keV annihilation photons with $\mathrm{TOT}_{\mathrm{hit}} \in [6, 7.25]$~ns$\cdot$V and prompt photons (one, two, or three) with $\mathrm{TOT}_{\mathrm{hit}} \in [8.5, 13.5]$~ns$\cdot$V. The hit multiplicity distribution for three classes of events is shown in Fig.~\ref{fig:selection_mn}(B), where the red-shaded region corresponds to the events with exactly three hits (two annihilation photons and one prompt photon), whereas  the blue-shaded region and pink-shaded region show the events with four and five hits, respectively.
For $^{55}$Co, events were analyzed for a hit multiplicity of three, consisting of two 511 keV annihilation photons with $\mathrm{TOT}_{\mathrm{hit}} \in [6, 7.8]$ ns$\cdot$V and one prompt photon with $\mathrm{TOT}_{\mathrm{hit}} \in [8.8, 13.5]$ ns$\cdot$V. All photons with $\mathrm{TOT}_{\mathrm{hit}}$ within this fixed prompt window were treated collectively as prompt, without resolving their energies. The corresponding hit multiplicity distribution ($\mu$), depicted in Fig.~\ref{fig:selection_co}(B), with the red-shaded region to highlight events with exactly two annihilation photons and one prompt photon, identified by their $\mathrm{TOT}_{\mathrm{hit}}$ values. 
The selection of annihilation and prompt candidates using the $\mathrm{TOT}_{\mathrm{hit}}$ windows, followed by event formation based on hit multiplicity, does not yield a fully pure data set. Photons that scatter in the phantom or detector material, prompt photons that deposit only part of their energy, and random coincidences can all be misclassified as either annihilation or prompt photons and thus contaminate the selected events. Only events within the red-shaded region of $\theta_{AA}$ in these panels were kept for further analysis. In the ideal case, the prompt gamma ($\gamma_P$) is emitted isotropically with respect to the annihilation photons. However,  the angular distribution between the annihilation photons and the prompt gamma shows an excess at small angles, which mainly originates from secondary scattered photons being identified as annihilation or prompt photons. To suppress such misidentified configurations and obtain a cleaner sample of true events, an additional requirement of $\theta_{PA} \geq 30^\circ$ was applied, as illustrated in Fig.~\ref{fig:selection_mn}(D) and Fig.~\ref{fig:selection_co}(D). 
As a final step, we used the space–time consistency of the detected hits to improve the purity of the events. For each selected event, we form pairs of hits and check whether the two signals could come from two different photons produced in the same annihilation, or whether one of them is more likely a secondary interaction of the other (intra-detector Compton scattering) or a random coincidence. This check is done with the so-called Scatter Test (ST). For a pair of hits with detected hit-times $t_j$ and $t_k$, with corresponding hit-positions $\vec{r}_j$ and $\vec{r}_k$, the ST is defined as: 
\begin{equation}
\text{ST} = |t_k - t_j| - |\vec{r}_k - \vec{r}_j|/c,
\end{equation}
where $c$ denotes the speed of light.
Under ideal conditions for direct photon detection ($|t_k - t_j| \leq |r_k - r_j|/c$), the ST yields a non-positive value. The test value approaches zero when both hits originate from the same photon that experiences two interactions (primary and secondary interactions). In contrast, ST becomes positive in the event of random coincidences, due to the expected substantial difference between $t_k$ and $t_j$. We analyze events with only pairs of annihilation hits with ST values $< -0.5$ ns, which significantly improved the overall purity of the events. The same criteria were adopted for candidate pairs of prompt photons (with four and five hit multiplicities) in the $^{52}$Mn data set to reduce detector-induced scattering effects. 
After applying the TOT-based selection, hit-multiplicity classification, angular cuts and the Scatter Test, we obtain a set of 
2$\gamma_a$ + $\gamma_p$ candidate events that can be used for further analysis. However, the purity of the events still depends on the detailed decay schemes of $^{52}$Mn and $^{55}$Co. Therefore, it is mandatory to identify and quantify the dominant background channels specific to these isotopes, discussed in the next section. 
\subsection*{Background contribution}\label{subsec4}
\subsubsection*{$^{52}$Mn}
Among the $\beta^+$+$\gamma$ emitters endorsed so far for PLI, $^{52}$Mn has several practical advantages. It has a half-life of about 5.6 days, which is long enough for production, transport, and multi-step experiments, but still suitable for preclinical use. In each decay ($^{52}\text{Mn} \rightarrow {}^{52}\text{Cr}^*$), there is a cascade of three prompt photons (744, 935 and 1434 keV) emitted almost simultaneously (within tens of picoseconds). 

\begin{figure}[htbp]
\centering
\includegraphics[width=0.87\linewidth]{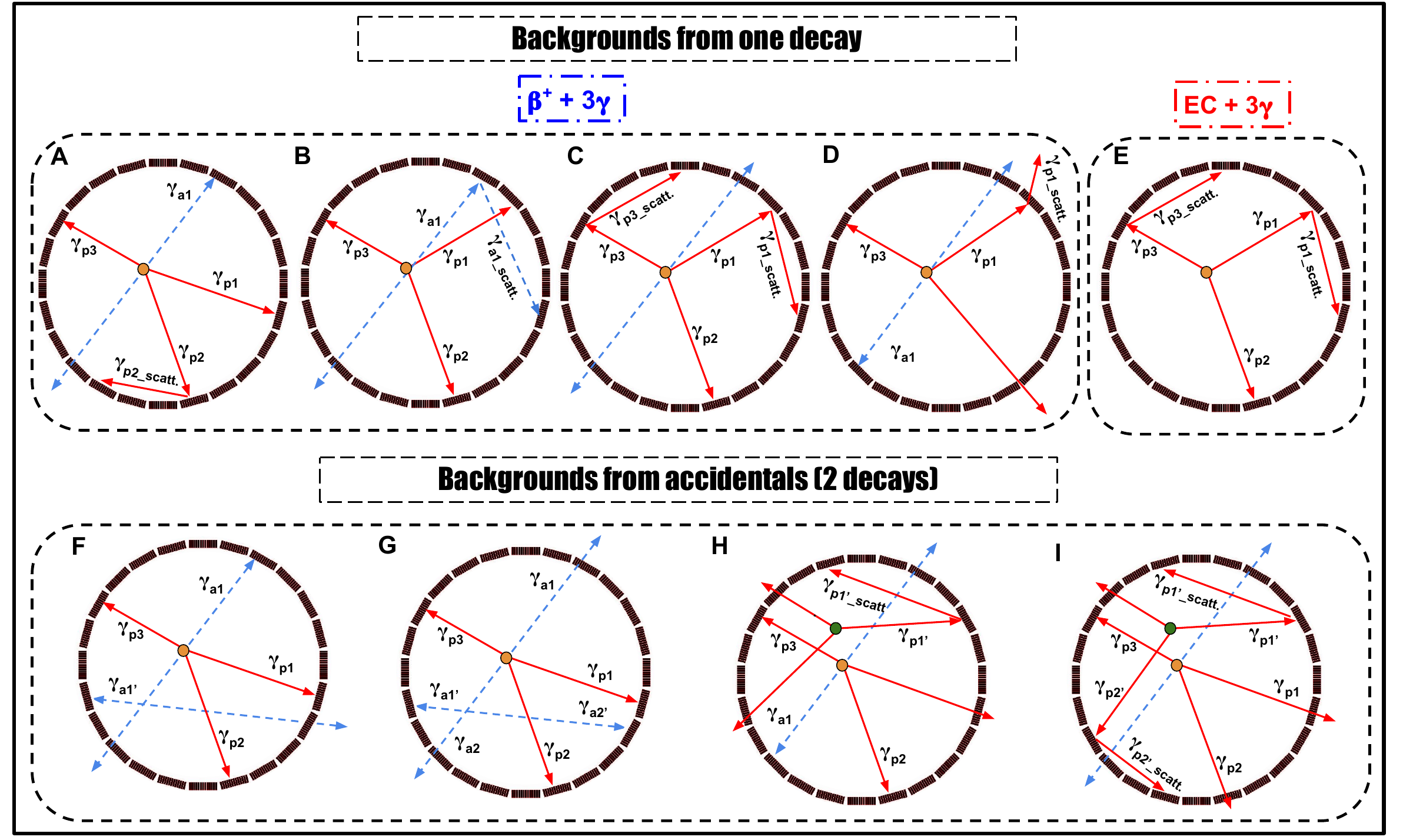}
\caption{ Cross section of the modular J-PET scanner showing background events for $^{52}$Mn. The primary and scattered prompt gamma is depicted as a red solid arrow, with primary and scattered annihilation photons as blue dashed arrows. \textbf{(A-E)} Background events from one decay: (A) One annihilation photon undetected, prompt gamma scatters twice, misidentified as an annihilation photon. (B) One annihilation photon undetected, the other scatters twice, misidentified as an annihilation photon. (C) Annihilation photon undetected, two prompt photons scatter, misidentified as an annihilation photon. (D) One annihilation photon undetected, one unidentified prompt gamma goes through low energy scattering misidentified as annihilation photon. (E) Electron capture without annihilation photons, two prompt photons scatter, misidentified as an annihilation photon.
\textbf{(F-I)} Background events from accidentals:
(F) Example of the background arising from the accidental coincidence where one annihilation photon from one event and one from another event. (G) Example of the background arising from the accidental coincidence of a prompt gamma from one event and annihilation photon from the other event. (H) Example of the background arising from the accidental coincidence of registering one or two prompt gammas and annihilation photon from one event and one prompt gamma and the prompt scattered registered as annihilation photon from the other event. (I) Example of the background arising from the accidental coincidence of registering two prompt gammas and the scattered prompt registered as annihilation photon and one prompt gamma from the other event. 
}
\label{fig:backroundMn}
\end{figure}

So, a single decay may provide up to three prompt candidates in coincidence with the annihilation photons. This increases the chance of forming usable  $2\gamma_a + \gamma_p$ events by roughly a factor of three compared to an isotope with only one prompt photon. However, a major drawback of using $^{52}$Mn is that in 70.6\% of decays these prompt photons are produced together with electron capture rather than positron emission. These decays do not contribute in useful events for PLI, rather, they significantly increase the scattering background. 
In our analysis, annihilation and prompt candidates are selected primarily on the basis of their $\mathrm{TOT}_{\mathrm{hit}}$ values. In the TOT spectra, both the 511~keV annihilation photons and the prompt photons from $^{52}$Mn give rise to pronounced structures that can be associated with the corresponding Compton edges. If the photon energies were well separated and only single prompt photons were present, one could, in principle, choose clean $\mathrm{TOT}_{\mathrm{hit}}$ windows around these regions and distinguish the two groups without much ambiguity. In plastic scintillators, however, photon interactions are dominated by Compton scattering, and high-energy prompt photons tend to scatter at small angles according to the Klein–Nishina formula. As a result, they often deposit only a fraction of their energy in a given strip, and the measured $\mathrm{TOT}_{\mathrm{hit}}$ can fall into the range expected for 511~keV photons, as seen in Fig.~\ref{fig:selection_mn}(A). The $^{52}$Mn cascade further enhances this effect and increases the number of prompt-related counts in the 511~keV region. This makes it more difficult to select high purity annihilation candidates. To reduce this contamination, we chose a relatively narrow $\mathrm{TOT}_{\mathrm{hit}}$ around the main 511~keV peak for annihilation photons. This improves the purity of the annihilation sample, but at the cost of statistics, partially counteracting the gain in efficiency expected from the three-fold prompt multiplicity. In addition to tightening the $\mathrm{TOT}_{\mathrm{hit}}$ window for 511~keV photons, we applied a sequence of further selection cuts based on angular correlation and ST to disentangle genuine annihilation pairs from prompt candidates in each event. Despite this, several types of background configurations can still bypass these selection criteria, as illustrated in~\ref{fig:backroundMn}. These background events can be divided mainly into two categories: contamination from a single decay (Fig.~\ref{fig:backroundMn}(A–E)) and accidental coincidences involving two decays (Fig.~\ref{fig:backroundMn}(F–I)).

In the first category, the background arises from single decays of $^{52}$Mn. The first scenario occurs in 29.4\% of cases, where prompt gamma photons are emitted after the positrons (Fig.~\ref{fig:backroundMn}(A–D)). A major fraction of these events can be eliminated by applying the ST between annihilation photon pairs and prompt photon pairs, combined with angular constraints (Fig.~\ref{fig:selection_mn}(D)). The second scenario occurs in 70.6\% of $^{52}$Mn decays, where prompt photons are emitted following electron capture rather than positron emission. Such events can be mitigated by applying the ST between pairs of prompt photons.
In the second class, the background arises from accidental coincidences involving two decays (Fig.~\ref{fig:backroundMn}(F–I)). Some of these events can be eliminated by restricting the angle range $\theta$ and applying ST.
In the analysis, we recorded a total of 9,323,325 events for $2\gamma_a + \gamma_p$ topology, 68,240 for $2\gamma_a + 2\gamma_p$, and 1,088 for $2\gamma_a + 3\gamma_p$. Thus, the $2\gamma_a + 2\gamma_p$ and $2\gamma_a + 3\gamma_p$ events together account for only 0.74\% of the $2\gamma_a + \gamma_p$ events, after all selection cuts. Consequently, 
these higher-multiplicity events were used only to study prompt multiplicity and were excluded from the positronium lifetime analysis, so they do not contribute to the background in the lifetime spectra.
\subsubsection*{$^{55}$Co} 
\begin{figure}[htbp]
\centering
\includegraphics[width=0.89\linewidth]{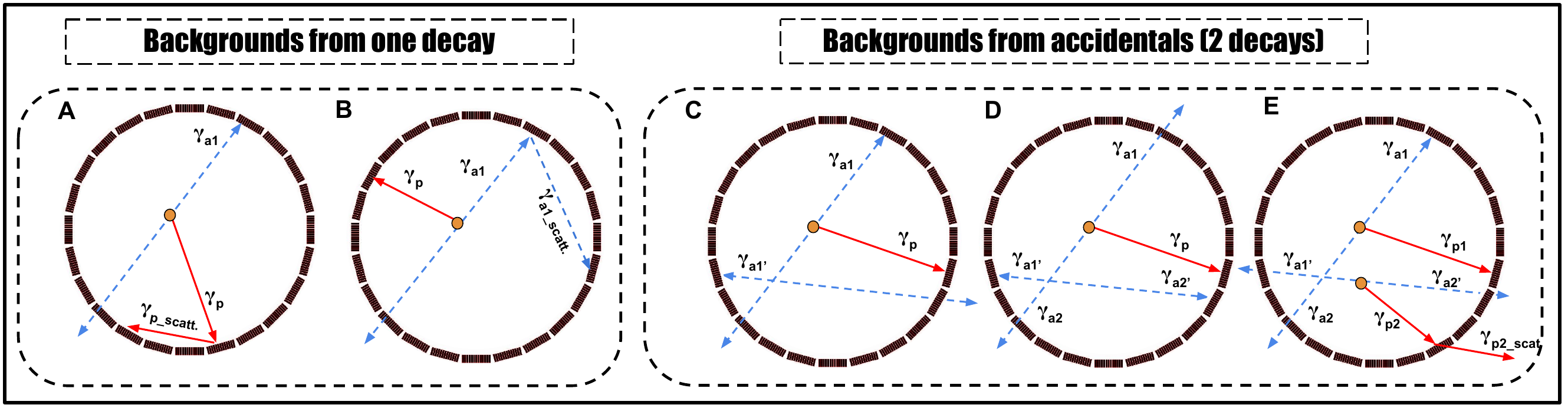}
\caption{ Cross-Section of the Modular J-PET Scanner: Background Events for $^{55}$Co The modular J-PET scanner illustrates background events for $^{55}$Co decays, with the prompt gamma depicted as a red solid arrow and primary/scattered annihilation photons as blue dashed arrows. \textbf{(A)} An example of a background event occurs when one annihilation photon goes undetected, and the prompt gamma scatters twice in the detector, leading to the misidentification of the scattered photon as an annihilation photon. \textbf{(B)} Another example involves one annihilation photon being undetected, with the other scattering twice within the detector, causing the scattered photon to be misidentified as an annihilation photon \textbf{(C)} A background event from an accidental coincidence arises when a prompt gamma from one decay event is detected alongside annihilation photons from another event \textbf{(D)} A background event from an accidental coincidence occurs when a prompt gamma and one annihilation photon from one decay event are registered with an annihilation photon from another event.
\textbf{(E)} An additional background event from an accidental coincidence occurs when a prompt gamma and one annihilation photon from one decay event are registered with an unidentified prompt gamma from another decay goes through low energy scattering misidentified as annihilation photon.
}
\label{fig:backroundCo}
\end{figure}

The $^{55}$Co isotope also has several features that are favorable for PLI. It has a half-life of 17.5~h, which is compatible with typical preclinical imaging including monoclonal antibodies and short enough to be considered for clinical applications, while still allowing for centralized production and shipment. About 76\% of $^{55}$Co decays proceed via $\beta^+$ emission, and in most of these decays a 931~keV $\gamma$ ray is emitted with high intensity, providing a well-defined prompt signal for building $2\gamma_a + \gamma_p$ events. The prompt scheme is therefore simpler than for $^{52}$Mn and naturally leads to a clean $2\gamma_a + \gamma_p$ topology. In addition, $^{55}$Co has already been used in preclinical PET studies with peptides and antibodies, indicating that suitable chelators and labelling strategies are available for biologically targeted imaging. However, in dosimetry and radiation safety, the presence of several additional $\gamma$ lines and the production of the long-lived $^{55}$Fe daughter must be taken into account when considering translational use. For PLI with plastic scintillators we also have to take into account the 477~keV photons from $^{55}$Co (intensity of order 15\%), which sits not far from the 511~keV annihilation energy. In the $\mathrm{TOT}_{\mathrm{hit}}$ spectra, photons from this transition often give values that fall into the same region as those from annihilation photons, as seen in Fig.~\ref{fig:selection_co}(A). Some 477~keV photons are therefore classified as annihilation candidates and add to the background in the annihilation window. In our analysis we treat the resulting $^{55}$Co background in two groups: events that come from a single decay [Fig.~\ref{fig:backroundCo}(A–B)] and events built from photons belonging to two different decays [Fig.~\ref{fig:backroundCo}(C–D)].

In the first class, the background arises from individual decays of $^{55}$Co. One scenario occurs when one of the annihilation photons is not detected and the 931~keV prompt photon undergoes two interactions in the detector. In this case, a scattered prompt hit can be mistakenly treated as an annihilation photon [Fig.~\ref{fig:backroundCo}(A)]. Such events tend to populate small values of the angle $\theta_{PA}$ between the prompt and annihilation candidates and can be partially suppressed by imposing a lower cut on $\theta_{PA}$, as shown in Fig.~\ref{fig:selection_co}(D). Another scenario is where one annihilation photon is lost and the other scatters twice, so that a scattered annihilation photon is misidentified as a second independent annihilation photon [Fig.~\ref{fig:backroundCo}(B)]; these configurations are reduced by applying the Scatter Test to the pair of annihilation candidates. In the second class, the background comes from accidental coincidences of photons from two separate $^{55}$Co decays. An example is when a prompt photon from one decay is detected together with an annihilation pair from another [Fig.~\ref{fig:backroundCo}(C)], where a fraction of such events can be rejected using the Scatter Test. Another example is when a prompt photon and one annihilation photon from one decay are detected together with an annihilation photon from a second decay [Fig.~\ref{fig:backroundCo}(D)]. In this case, restricting the allowed range of the opening angle $\theta$ [Fig.~\ref{fig:selection_co}(C–D)], in combination with the Scatter Test, further suppresses these accidental configurations.

\subsection*{Positronium Lifetime estimation}\label{subsec5}

After applying all selection criteria described above, we obtained a purified sample of 2$\gamma_a$ + $\gamma_p$ events, which were used for the positronium lifetime reconstruction. For each selected $2\gamma_a + \gamma_p$ event, the annihilation point ($\vec{r}_a$) and annihilation time ($t_a$) are determined from the times and positions of the annihilation photon hits, denoted as $(t_1, \vec{r}_1)$ and $(t_2, \vec{r}_2)$:
\begin{equation}\label{eq:posRecoFormula}
\vec{r}_a = \frac{\vec{r}_1 + \vec{r}_2}{2} 
        + \frac{c\left(t_2 - t_1\right)}{2} 
          \cdot \frac{\vec{r}_1 - \vec{r}_2}{|\vec{r}_1 - \vec{r}_2|},
\end{equation}
and
\begin{equation}
t_{a} = \frac{t_1 + t_2}{2} - \frac{|\vec{r}_1 - \vec{r}_2|}{2c}.
\end{equation}
The first term in Eq.~\ref{eq:posRecoFormula} gives the midpoint between the two hit positions, while the second term shifts this midpoint along the line of response according to the measured time difference ($t_2 - t_1$). In this way, the annihilation point is obtained on an event-by-event basis without any external timing reference. Analogously, the annihilation time $t_{a}$ is calculated as the average of the two detection times, corrected by half of the photon flight time between the hit positions. It should be mentioned that the annihilation point distribution was obtained using only the annihilation photon pair (511~keV), while the additional prompt gamma photon was used to estimate the lifetime of the positron. The positron lifetime is defined as
\[
\Delta T = t_a - t_p,
\]
i.e. the time interval between positron emission and annihilation. Here, $t_p$ is used as an estimator of the positron emission time. According to the decay schemes in Fig.~\ref{fig:decay_scheme}(B–C), the prompt gamma rays for both isotopes are emitted, on average, within about 50~ps of the positron emission, which is much shorter positronium lifetimes and detector time resolution. It is therefore a good approximation to identify the prompt emission time with the positron emission time. The emission time $t_p$ is obtained by correcting the registered prompt time for the photon time of flight:
\begin{equation} 
t_p =  t_i -  \frac{|\vec{r}_i - \vec{r}_a|}{c}, 
\end{equation}
where $t_i$ and $\vec{r}_i$ are the detection time and position of the prompt gamma, and $\vec{r}_a$ is the annihilation point reconstructed using Eq.~\ref{eq:posRecoFormula}. The positrons emitted in the decays of $^{52}$Mn and $^{55}$Co have very short ranges in the tissue, approximately 0.58~mm and 1.72~mm, respectively~\cite{ManishBAMS, Moskal2025IEEE}. This means, one can safely assume that the creation point of the positron and the annihilation site are separated by less than a few millimeters. When combined with the sub-nanosecond timing, this justifies treating the prompt emission point as effectively co-located with the annihilation site for the purpose of constructing positron lifetime spectra ($\Delta T$). The obtained spectra
contain contributions from the direct annihilation of positrons as well as their annihilation through the formation of Ps (pPs, oPs) atoms. 
The $\Delta T$ spectrum was analyzed by fitting a sum of exponential decay components convoluted with detector resolution, using the specialized PALS Avalanche software to determine the mean oPs lifetime, as described in \cite{doi:10.1126/sciadv.abh4394, PALSAva1, PALSAva2, PALSAva3}. The value of the initial parameters for para-positronium (pPs) and direct annihilation was set at 125 ps (with a relative intensity of 10\%) and 388 ps with an intensity of 60\%, respectively. Both lifetime and intensities of these two parameters were restricted to vary up to twice the initial set values. For CRM measurements, an additional fixed component corresponding to Parafilm was included, with a lifetime of 2.35~ns and an intensity of 10\%, reflecting the known contribution of the Parafilm layers used in the sample mounting. For tissue measurements, no additional components were added for the Eppendorf or PCR tubes, since their contribution to the overall annihilation statistics was verified to be negligible. The oPs component was left unconstrained in both lifetime and intensity, as its determination is the main aim of the analysis. A constant background level was estimated by averaging the counts in the range of $-10$~to $-5$~ns and subtracted before fitting, so that the fit is driven by the physical decay components. 

In addition to the mean oPs lifetime $\tau_\text{oPs}$ obtained from the multi-component fit, we also evaluated the mean positron lifetime $\Delta T_{\text{mean}}$, defined as the average value of $\Delta T$ between 0~ns and 5~ns after background subtraction, as described in~\cite{doi:10.1126/sciadv.adp2840, ManishIEEE}. This quantity is sensitive to the overall shape of the lifetime spectrum, including both short- and long-lived components, and provides a complementary measure to $\tau_\text{oPs}$. The extracted values of $\tau_\text{oPs}$ and $\Delta T_{\text{mean}}$ for all samples are presented and discussed in the Results section.
\section*{Results}

In this section, we evaluate the performance of $^{52}$Mn and $^{55}$Co as radionuclides for PLI with the modular J-PET. The annihilation point distribution ($\vec{r}_a$) [Eq.~\ref{eq:posRecoFormula}] was obtained from $2\gamma_a + \gamma_p$ events built from 511 keV photon pairs, and the resulting images for $^{52}$Mn and $^{55}$Co are shown in Fig.~\ref{fig:result_mn}(A) and Fig.~\ref{fig:result_co}(A), respectively. The images were reconstructed on a uniform voxel grid with size 2.5 mm in each direction. For $^{52}$Mn, 9,323,325 $2\gamma_a + \gamma_p$ events passed the selection criteria, whereas for $^{55}$Co the corresponding number was 3,706,231. To characterize signal quality, the peak-to-background ratio (PBR) was calculated for each isotope, yielding PBR = 87 for $^{52}$Mn and PBR = 133 for $^{55}$Co. The higher PBR value observed for $^{55}$Co indicates a substantially cleaner $2\gamma_a + \gamma_p$ selection, reflecting the simpler decay scheme and reduced background contribution due to  electron capture induced prompt photons. For quantitative PLI analysis, a spherical region of interest (ROI) of radius 33 mm was defined at the centre of each sample. The average number of events across the four samples was approximately 239760 for $^{52}$Mn and 179450 for $^{55}$Co. The resulting positron lifetime spectra $\Delta T$ for all four samples are shown in Fig. 7(D) for $^{52}$Mn and Fig. 8(D) for $^{55}$Co.
\begin{table*}[htbp]
\caption{Positronium lifetime and intensity obtained from fitting the positron lifetime spectra.
$\tau_\text{oPs}$ is the mean lifetime of oPs. $I_\text{oPs}$, $I_\text{direct}$, and $I_\text{pPs}$
are the relative intensities corresponding to oPs annihilation, direct positron-electron annihilation,
and pPs annihilation, respectively. $\Delta T_\mathrm{mean}$ denotes the mean positron lifetime. All uncertainties are given at the $\pm1\sigma$ level. Note that the $\tau_\text{oPs}$ value for adipose and myxoma tissues are different than previously published~\cite{EJNMMI2023, doi:10.1126/sciadv.abh4394} because they were stored in refrigerator for few months (three months in case of $^{52}$Mn and five months in case of $^{55}$Co) and underwent chemical degradations.
}
\centering
\small
\begin{tabular*}{\textwidth}{@{\extracolsep{\fill}} c c c c c c @{\extracolsep{\fill}}}
\toprule

\multicolumn{6}{c}{$^{52}$Mn} \\
\midrule
{Sample Name} & {$\tau_\text{oPs}$ [ns]} & {$I_\text{oPs}$ [\%]} & {$I_\text{direct}$ [\%]} & {$I_\text{pPs}$ [\%]} & {$\Delta T_\text{mean}$ [ns]} \\
\midrule
Polycarbonate & 2.069 $\pm$ 0.020 & 21.44 $\pm$ 0.17 & 54.82 $\pm$ 0.21 & 13.74 $\pm$ 0.16 & 1.120 $\pm$ 0.002 \\
Fused Silica  & 1.836 $\pm$ 0.013 & 32.45 $\pm$ 0.20 & 49.46 $\pm$ 0.20 & 8.10  $\pm$ 0.16 & 1.223 $\pm$ 0.003 \\
Myxoma (see caption)       & 2.111 $\pm$ 0.018 & 24.15 $\pm$ 0.17 & 64.07 $\pm$ 0.22 & 11.78 $\pm$ 0.18 & 1.043 $\pm$ 0.002 \\
Adipose (see caption) & 2.081 $\pm$ 0.018 & 23.68 $\pm$ 0.17 & 63.41 $\pm$ 0.22 & 12.91 $\pm$ 0.19 & 1.044 $\pm$ 0.002 \\
\midrule
\multicolumn{6}{c}{$^{55}$Co} \\
\midrule
{Sample Name} & {$\tau_\text{oPs}$ [ns]} & {$I_\text{oPs}$ [\%]} & {$I_\text{direct}$ [\%]} & {$I_\text{pPs}$ [\%]} & {$\Delta T_\text{mean}$ [ns]} \\
\midrule
Polycarbonate & 2.174 $\pm$ 0.024 & 20.34 $\pm$ 0.21 & 59.76 $\pm$ 0.24 & 9.90 $\pm$ 0.19 & 1.136 $\pm$ 0.002 \\
Fused Silica  & 1.607 $\pm$ 0.012 & 36.30 $\pm$ 0.22 & 42.54 $\pm$ 0.22 & 11.15 $\pm$ 0.18 & 1.240 $\pm$ 0.002 \\
Myxoma (see caption) & 2.070 $\pm$ 0.029 & 13.46 $\pm$ 0.49 & 72.15 $\pm$ 2.43 & 14.39 $\pm$ 0.79 & 0.983 $\pm$ 0.002 \\
Adipose (see caption)& 1.996 $\pm$ 0.030 & 11.98 $\pm$ 0.92 & 71.87 $\pm$ 0.16 & 16.15 $\pm$ 0.13 & 0.972 $\pm$ 0.002 \\
\midrule

\end{tabular*}
\label{table:fit}
\end{table*}
The lifetime spectra were fitted with a multi-component model including direct annihilation, para-positronium (pPs) and ortho-positronium (oPs) components, convolved with the detector time response as described previously. The fitted mean oPs lifetimes $\tau_{\text{oPs}}$, the corresponding intensities $I_{\text{oPs}}$, and the intensities of the direct and pPs components ($I_{\text{direct}}$, $I_{\text{pPs}}$) are summarized in Table 1 together with the mean positron lifetime $\Delta T_{\text{mean}}$ obtained as the average of $\Delta T$ between 0 and 5~ns.
\subsection*{Performance on Certified Reference Materials}

\begin{figure}[htbp]
\centering
\includegraphics[width=0.95\linewidth]{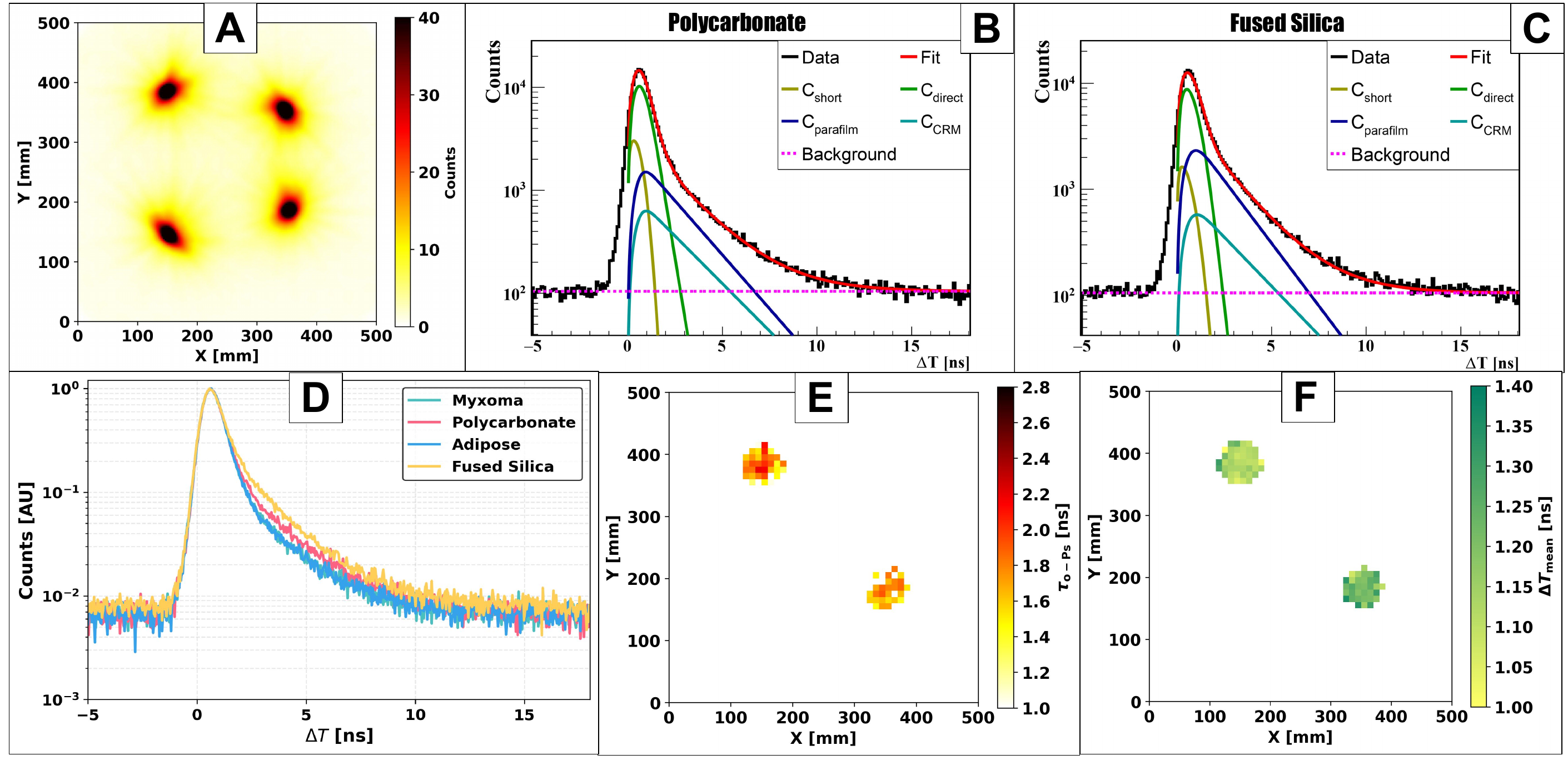}
\caption{Results for $^{52}$Mn: {\bf{(A)}} Transaxial view of the annihilation point distribution ($\vec{r}_a$) for $2\gamma_a + \gamma_p$ events. {\bf{(B-C)}} Distributions of positron annihilation lifetimes ($\Delta T$) for the Polycarbonate (B) and Silica (C). The black histograms represent the experimental data, while the overlaid curves correspond to the fitted components: pPs (C$_\text{short}$), direct annihilations (C$_\text{direct}$), oPs contribution from parafilm(C$_\text{parafilm}$), oPs contribution from CRM (C$_\text{CRM}$), and background from accidental coincidences. The red curve represents the total fit, obtained as the sum of all contributions. {\bf{(D)}}  Distributions of positron annihilation lifetimes ($\Delta T$) for Myxoma, Polycarbonate, Adipose and Fused silica. {\bf{(E-F)}} Mean oPs lifetime ($\tau_\text{oPs}$) (E) and mean positron lifetime ($\Delta T_\mathrm{mean}$) (F) in the Polycabonate and Silica. }

\label{fig:result_mn}
\end{figure}

\begin{figure}[htbp]
\centering
\includegraphics[width=0.95\linewidth]{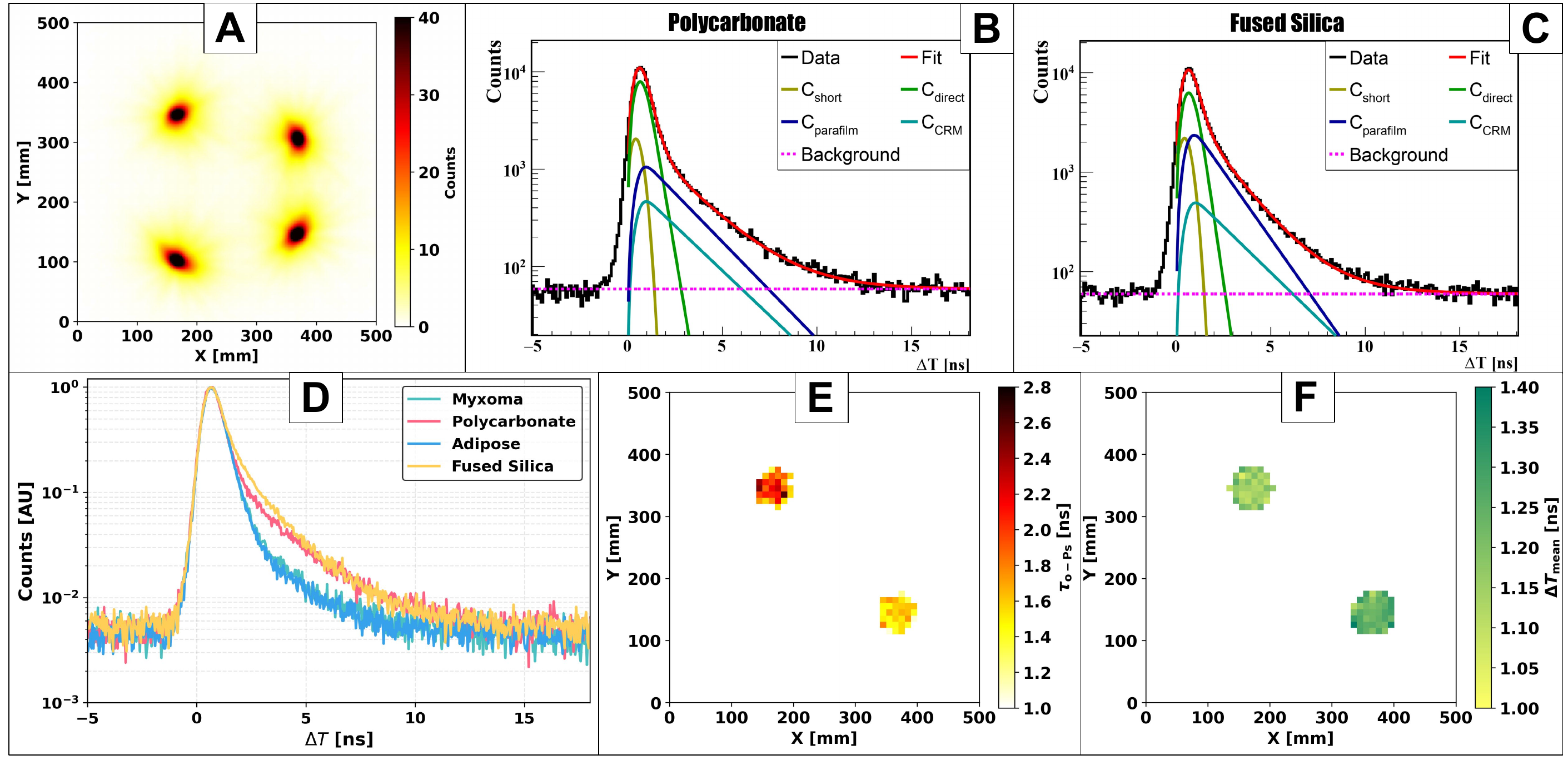}
\caption{Results for $^{55}$Co: {\bf{(A)}} Transaxial view of the annihilation point distribution ($\vec{r}_a$) for $2\gamma_a + \gamma_p$ events. {\bf{(B-C)}} Distributions of positron annihilation lifetimes ($\Delta T$) for the Polycarbonate (B) and Silica (C). The black histograms represent the experimental data, while the overlaid curves correspond to the fitted components: pPs (C$_\text{short}$), direct annihilations (C$_\text{direct}$), oPs contribution from parafilm(C$_\text{parafilm}$), oPs contribution from CRM (C$_\text{CRM}$), and background from accidental coincidences. The red curve represents the total fit, obtained as the sum of all contributions. {\bf{(D)}}  Distributions of positron annihilation lifetimes ($\Delta T$) for Myxoma, Polycarbonate, Adipose and Fused silica. {\bf{(E-F)}} Mean oPs lifetime ($\tau_\text{oPs}$) (E) and mean positron lifetime ($\Delta T_\mathrm{mean}$) (F) in the Polycabonate and Silica }

\label{fig:result_co}
\end{figure}

The CRMs from AIST (polycarbonate and fused silica) were used as benchmarks to validate the PLI methodology. The certified oPs lifetimes are $2.10 \pm 0.05$~ns for polycarbonate and $1.62 \pm 0.05$~ns for fused silica. For the polycarbonate material, both isotopes produce results that match the certified value within their respective uncertainty levels. The $^{52}$Mn and $^{55}$Co isotopes generate almost the same $\tau_{\text{oPs}}$ values of $2.069 \pm 0.020$~ns and $2.174 \pm 0.024$~ns respectively (Table~1, Fig.~\ref{fig:result_mn}(B), Fig.~\ref{fig:result_co}(B)). The $^{55}$Co measurement of $\tau_{\text{oPs}}$ for fused silica matches the certified value at $1.607 \pm 0.012$~ns but the $^{52}$Mn measurement deviates by 200~ps from the official standard (Fig.~\ref{fig:result_mn}(C), Fig.~\ref{fig:result_co}(C)).
In all CRM measurements, ten layers of Parafilm (nominal total thickness $\approx 1.2$~mm) were introduced between the plates to host the liquid source. The Parafilm contribution was modelled as an additional oPs component with fixed lifetime 2.35~ns and fixed intensity 10\%. The $^{52}$Mn fused silica data show $\tau_{\text{oPs}}$ values closer to the certified silica lifetime when the Parafilm intensity is allowed to change, which indicates a local thickness or positioning problem in the Parafilm spacer for this specific setup. Since the $^{52}$Mn and $^{55}$Co measurements were performed at different times, small differences in the effective Parafilm thickness cannot be excluded. In our final analysis, we therefore applied a constant 10\% Parafilm intensity to all data sets for consistency and treated the remaining deviation as a systematic effect.
To further cross-check the spectral decomposition, we evaluated the ratio of oPs intensities between polycarbonate and fused silica. An interlaboratory comparison involving 12 laboratories reported $I_{\text{poly}}/I_{\text{silica}} \approx 0.56$ for these CRMs~\cite{CRMInte}. For $^{55}$Co we obtain $I_{\text{poly}}/I_{\text{silica}} = 0.560 \pm 0.006$, in excellent agreement with the interlaboratory value, while for $^{52}$Mn we find a slightly larger ratio of $0.661 \pm 0.007$, consistent with the aforementioned Parafilm-related bias in the fused silica data.
Despite these differences in $\tau_{\text{oPs}}$, the mean positron lifetimes $\Delta T_{\text{mean}}$ for polycarbonate and fused silica remain very stable between isotopes. For both CRMs, the $\Delta T_{\text{mean}}$ values measured with $^{52}$Mn and $^{55}$Co differ by less than 20~ps (Table~1). This confirms that $\Delta T_{\text{mean}}$, which integrates information over all annihilation channels, is a robust and isotope-independent PLI observable~\cite{doi:10.1126/sciadv.abh4394, ManishIEEE}.

\subsection*{Tissue samples and voxel-wise PLI estimation}

For the cardiac myxoma and adipose tissue samples, the $\tau_{\text{oPs}}  $ values obtained with both isotopes differ from the \textit{ex-vivo} values reported previously for freshly used tissues  ($\tau_{\text{oPs}} \approx 1.9$~ns for myxoma and $\approx 2.6$–$2.7$~ns for adipose)~\cite{EJNMMI2023, doi:10.1126/sciadv.abh4394}. In the present measurements, $\tau_{\text{oPs}}$ is reduced to $2.111 \pm 0.018$~ns (myxoma) and $2.081 \pm 0.018$~ns (adipose) for $^{52}$Mn, and to $2.070 \pm 0.029$~ns (myxoma) and $1.996 \pm 0.030$~ns (adipose) for $^{55}$Co (Table~1). The corresponding $\Delta T_{\text{mean}}$ values show smaller differences between isotopes: the $^{52}$Mn and $^{55}$Co measurements differ by 60~ps (myxoma) and 72~ps (adipose), respectively. These shifts indicate that the long-term storage protocol (five years in formaldehyde followed by staining with Lugol's solution and refrigerated storage for several months) altered the microscopic structure of the tissues, an effect that is further discussed below.
Interestingly, the oPs intensities in tissues are systematically higher for $^{52}$Mn than for $^{55}$Co (Table~1). The myxoma and adipose samples stored for three months after staining ($^{52}$Mn measurement) exhibit higher $I_{\text{oPs}}$ than the same samples measured with $^{55}$Co after five months of storage, supporting the hypothesis of progressive microstructural deterioration over time.
To visualize the spatial distribution of PLI parameters, voxel-wise oPs mean lifetimes and mean positron lifetimes were reconstructed. In the transverse plane, $\tau_{\text{oPs}}$ maps were computed with $1$~cm~$\times$~$1$~cm voxels, integrating over the full axial extent of the scanner (Fig.~7(E) and Fig.~8(E)). Voxel-wise $\Delta T_{\text{mean}}$ maps were obtained with 1~cm isotropic voxels (Fig.~7(F) and Fig.~8(F)). In both isotopes, $\tau_{\text{oPs}}$ is systematically lower in fused silica than in polycarbonate, while $\Delta T_{\text{mean}}$ is higher in fused silica. This pattern reflects the higher oPs intensity in fused silica, which shifts the effective $\Delta T_{\text{mean}}$ to longer values even though the oPs lifetime itself is shorter. The voxel-wise reconstructions confirm that both $^{52}$Mn and $^{55}$Co enable reliable PLI of heterogeneous samples with the modular J-PET system.
However, for cardiac myxoma and adipose tissue samples, we did not reconstruct voxel maps of $\tau_{\text{oPs}}$ and $\Delta T_{\text{mean}}$. Instead, only ROI-averaged values are reported in Table~1, since the small sample size and limited statistics (also affected by long-term storage) did not allow for a stable voxel-wise analysis.
%
%%%%%%%%%%%%%%%%%%%%%%%%---------------------------------------
\section*{Discussion}

In this work, we performed the first PLI studies with the $\beta^+ + \gamma$ emitters $^{52}$Mn and $^{55}$Co using the modular J-PET scanner. We selected $2\gamma_a + \gamma_p$ events for both radionuclides, obtaining about $9.3\times10^6$ events for $^{52}$Mn and $3.7\times10^6$ events for $^{55}$Co after all selection criteria were applied. These data sets are large enough to assess how well each isotope performs for PLI in a realistic multi-sample arrangement. For the validation of lifetime analysis, CRM (polycarbonate and fused silica) were used as the benchmarks. For the certified reference materials, we first checked whether the oPs lifetimes extracted from the fits are consistent with the values provided by the manufacturers. In polycarbonate, the $\tau_{\text{oPs}}$ obtained with $^{52}$Mn and $^{55}$Co, $2.069 \pm 0.020$~ns and $2.174 \pm 0.024$~ns, both lie within the quoted uncertainty of the certified lifetime and are close to each other. This suggests that, at least for this material, the selection of $2\gamma_a + \gamma_p$ events and subsequent lifetime analysis do not strongly depend on which of the two isotopes is used. However, the results for fused-silica show a different pattern. Here, the $^{55}$Co result ($1.607 \pm 0.012$~ns) reproduces the certified value, while the $^{52}$Mn measurement is higher by about 200~ps. Since the Parafilm spacer was included in the fit as a fixed oPs component, a natural explanation is that the $^{52}$Mn data are more sensitive to small differences in the effective Parafilm thickness or positioning in that particular configuration. This interpretation is compatible with the observed ratio $I_{\text{poly}}/I_{\text{silica}}$~\cite{CRMInte}, which for $^{55}$Co agrees with the interlaboratory value, whereas for $^{52}$Mn it is noticeably larger. In the present analysis we therefore treat the fused silica shift for $^{52}$Mn as a systematic effect linked to the mounting procedure rather than as a limitation of the isotope itself for PLI.

In contrast, tissue measurements are clearly influenced by the extended storage of the samples. Both myxoma and adipose tissue show $\tau_{\text{oPs}}$ values that are substantially shorter than previously reported for freshly used tissues in \textit{ex-vivo} PLI with $^{22}$Na~\cite{EJNMMI2023,doi:10.1126/sciadv.abh4394}. This difference most likely resulted from chemical degradation of the tissue due to prolonged (five year) storage, which could have caused deleterious effect resulting in structural changes in macromolecules and oxidation of lipids that make up adipose tissue~\cite{Likhithaswamy2022}. Interestingly, it was recently observed that paraformaldehyde “aging” causes deterioration of the chemical quality of biological samples, which can be analysed by means of spectral (ATR-FTIR) and spectroscopic (ToF-SIMS) methods~\cite{Moyo2025for}. Additionally, sample contrasting may also result in changes in the nanostructure of macromolecules. The iodine used in Lugol's solution binds to carbohydrates and glycolipids. It is not possible to assess to what extent sample contrasting may have influenced the structural changes in tissue macromolecules. What we observed — a decrease in $\tau_{\text{oPs}}$ — shows that this parameter is very sensitive, revealing structural changes resulting from the aging of biological material and its potential biodegradation. It is very likely that tissue shrinkage caused a condensation of volume, which may have altered the proportion between free voids and matter, for which $I_{\text{oPs}}$ serves as an indicator. The fact that $\tau_{\text{oPs}}$ and $I_{\text{oPs}}$ are systematically lower for the $^{55}$Co data, acquired later (about five months after drying) than the $^{52}$Mn data (about three months after drying), is compatible with progressive microstructural degradation during subsequent storage. The corresponding shifts in $\Delta T_{\text{mean}}$ (60–72~ps) point in the same direction. Due to the limited size of the tissue samples and the reduced statistics after all selection criteria, we restricted the analysis to ROI-averaged values and did not attempt voxel-wise $\tau_{\text{oPs}}$ and $\Delta T_{\text{mean}}$ maps for myxoma and adipose tissue. A dedicated study of storage induced changes in PLI parameters based on systematically prepared tissue samples will be needed before any quantitative biological conclusions can be drawn from such long-stored specimens. In the present work, tissue data are primarily used to compare isotope performance under realistic conditions. On this basis, we compared the performance of $^{52}$Mn and $^{55}$Co in terms of the lifetime analysis, since this is essential to assess their suitability as tracers for PLI. A higher peak-to-background ratio is obtained for $^{55}$Co ($\approx 133$) than for $^{52}$Mn ($\approx 87$), due to the simpler decay scheme of $^{55}$Co with a single dominant 931~keV prompt line. In contrast, $^{52}$Mn offers a three-photon prompt cascade, which in principle could enhance the efficiency of prompt detection by up to a factor of three. However, the large electron-capture fraction and the resulting increase in background from scattered prompt photons largely offset this advantage. The need for tight $\mathrm{TOT}_{\mathrm{hit}}$ windows of annihilation photons, together with additional angular and Scatter-Test cuts, further reduces the effective statistics. Therefore, we can conclude that, under identical acquisition and analysis conditions, $^{55}$Co achieves higher PBR and somewhat cleaner lifetime spectra for PLI.

Recently, several groups reported feasibility studies using $^{44}$Sc for PLI applications. Its physical properties are very close to what is desirable for a $\beta^+ + \gamma$ isotope: a half-life of 4.04~h, a high $\beta^+$ branching ratio of 94.3\%, and an almost always present 1157~keV prompt photon with negligible delay. Our results show that $^{55}$Co can complement the use of $^{44}$Sc when cobalt-based tracers or delayed imaging protocols are of interest. The more complex decay pattern of $^{52}$Mn introduces additional background, but it is likely to be better suited to applications where its long half-life or specific manganese chemistry are required. According to current regulatory, $^{52}$Mn and $^{55}$Co are not yet included among the radionuclides used in routine clinical PET. However, several of their properties make them attractive candidates for further development. For $^{55}$Co, the 17.5-hour half-life offers a significant advantage over shorter-lived isotopes such as $^{68}$Ga~\cite{Co_braad}. 
The longer half-life enables to application in immuno-PET imaging using monoclonal antibodies~\cite{Nayak2009} and to conduct delayed imaging, which improves tumor-to-background ratios, particularly in cancer studies. Preclinical studies using  $^{55}$Co-based tracers has shown promising results~\cite{1Co2017, Houson2022Co}. Beyond oncology, $^{55}$Co has been explored as a PET tracer for assessing neuronal damage in stroke patients~\cite{Stevens199955Cobalt}. Furthermore, multiplexed PET imaging using $^{55}$Co and $^{18}$F-labeled tracers in a single session has successfully separated their signals by leveraging the prompt-gamma emission of $^{55}$Co for triple coincidence detection~\cite{Co_multi, Co_multi2}. In contrast, $^{52}$Mn has been investigated in a range of applications. It has been used for \textit{in-vivo} immuno-PET imaging~\cite{Graves2015Mnimmu}. Owing to its properties, $^{52}$Mn is also suited for combined PET/MR imaging or as a PET analog of Mn-based MRI contrast agents to study their \textit{in-vivo} distribution and pharmacokinetics~\cite{Saar2018Mn, Zhou2020Mn, Lewis2015Mn, Vanasschen2016Mn}. Moreover, $^{52}$Mn can serve as a surrogate for Ca$^{2+}$ uptake studies, for example in the context of diabetes~\cite{Hernandez2017Mndia}. $^{52}$Mn-labeled liposomes have further been evaluated \textit{in-vivo}~\cite{Jensen2018Mn2}. In addition, the radiation effective dose of $^{52}$Mn administered as chloride to humans has been determined~\cite{DeNardo2019Mn3}, and its radiation toxicity has been evaluated in rats~\cite{Napieczynska2017Mn4}.

More importantly, the analysis strategy presented in this work shows that the J-PET methodology (event selection, Scatter-Test, use of $\tau_{\text{oPs}}$ and $\Delta T_{\text{mean}}$) can be applied to radionuclides with more complicated decay schemes, which is relevant for ongoing PLI activity on different scanner types. Moreover, the cascade emission of gamma photons by $^{55}$Co and $^{52}$Mn will enable studies of the correlations between these photons as discussed recently e.g. in reference~\cite{SHIMAZOE2023, Shimazoe2022, ShimazoeBAMS2022, Shimazoe2bams2022}.

%
%%%%%%
\section*{Conclusion}

This study reports the first successful experimental demonstration of PLI using the isotopes $^{52}$Mn and $^{55}$Co. Using the common event-selection scheme, we assessed their performance for the first time in a multi-sample configuration with the modular J-PET scanner. For both isotopes $^{52}$Mn and $^{55}$Co, the adopted methodology reproduced the lifetime values of the CRM materials provided by the manufacturer. The results show the reliability of reconstructing lifetimes and background estimation with J-PET even when working with radionuclides having complex decay patterns. Measurements with long-stored tissue samples (cardiac myxoma and adipose tissue) were used mainly to examine how the isotopes behave under realistic conditions.  The changes seen in the positronium parameters in these samples are most likely affected by the long fixation and storage, and should not be interpreted straightforwardly as reflecting only biological differences. The measurement of $^{55}$Co results in better effective PBR values and produces spectra with improved lifetime resolution. The longer half-life of $^{52}$Mn makes it suitable for manganese-based tracer applications, yet it faces contamination risks due to its multi-prompt photons scattered background. In comparison with earlier results for $^{44}$Sc, $^{44}$Sc and $^{55}$Co are favourable options for PLI, whereas $^{52}$Mn can be used for a more specialised situations linked to its radiochemistry and half-life. Combined with further measurements using the modular J-PET and clinical long-axial-FOV scanners, such work should enable a more quantitative comparison of candidate radionuclides and help separate the effects of isotope choice, scanner design and sample preparation on the observed PLI contrast. In parallel, the planned total-body J-PET system, with its expected gain in sensitivity over the present prototype, together with continuing progress in large axial field-of-view clinical scanners~\cite{Moskal-NEMA-PMB-2021,doi:10.1126/sciadv.adp2840,Steinberger2024,Prenosil-quadra-2022,Spencer-uExplorer-NEMA-2021, Li2024} and dedicated PLI reconstruction methods~\cite{Huang2025Fast,Huang2024High,Huang2025Statistical,Qi-positronium-2023,Berens2024analytic,Qi-positronium-2022,Chen2023,Shopa-Kamil-Bams2023,Huang_mic,Huangjnumed.125.270130,Chen2024}, should further improve the practical sensitivity and image quality achievable for PLI studies with $^{44}$Sc, $^{55}$Co and $^{52}$Mn. 
\section*{Data availability}
The datasets collected in the experiment and analyzed during the
current study are available under restricted access due to the large data volume. Direct access to the data can be arranged on request by contacting the corresponding author.

\section*{Funding}

This study is supported by the National Science Centre of Poland through grants MAESTRO no. 2021/42/A/ST2/00423 (P.M.), OPUS no. 2021/43/B/ST2/02150 (P.M.), OPUS24+LAP no. 2022/47/I/NZ7/03112 (E.Ł.S.) and SONATA no. 2023/50/E/ST2/00574 (S.S.), the Ministry of Science and Higher Education through grant no. IAL/SP/596235/2023 (P.M.) and SPUB/SP/627733/2025 (E.Ł.S.), the SciMat and qLife Priority Research Areas budget under the program Excellence Initiative – Research University at Jagiellonian University (P.M. and E.Ł.S.), the Research Support Module as part of the Excellence Initiative – Research University program at Jagiellonian University (M.D.), European Union within the Horizon Europe Framework Programme (ERC Advanced Grant POSITRONIUM no. 101199807) and PRISMAP via Project\_1729020993\_aniX3. We also acknowledge Polish high-performance computing infrastructure PLGrid (HPC Center: ACK Cyfronet AGH) for providing computer facilities and support within computational grant no. PLG/2024/017688 and PLG/2025/018762.

\bibliography{sample}

@article{Triffitt1970boneuptake,
	journal = {Calcified Tissue Research},
	doi = {10.1007/bf02196185},
	issn = {0008-0594},
	number = {1},
	language = {en},
	publisher = {Springer Science and Business Media LLC},
	title = "{The uptake of sodium-22 by bone and certain soft tissues}",
	url = {http://dx.doi.org/10.1007/BF02196185},
	volume = {6},
	author = {Triffitt, J. T. and Neuman, W. F.},
	pages = {70--76},
	date = {1970-12},
	year = {1970},
	month = {Dec.},
}

@article{Chen2024,
	journal = {Frontiers in Physics},
	doi = {10.3389/fphy.2024.1429344},
	issn = {2296-424X},
	publisher = {Frontiers Media SA},
	title = "{Enhanced positronium lifetime imaging through two-component reconstruction in time-of-flight positron emission tomography}",
	url = {http://dx.doi.org/10.3389/fphy.2024.1429344},
	volume = {12},
	author = {Chen, Zhuo and Kao, Chien-Min and Huang, Hsiun-Hsiung and An, Lingling},
	date = {2024-07-15},
    pages = {1429344},
	year = {2024},
	month = {Jul.},
	day = {15},
}

@article{Chen2023,
	journal = {Bio-Algorithms and Med-Systems},
	doi = {10.5604/01.3001.0054.1807},
	issn = {1896-530X},
	number = {1},
	publisher = {Index Copernicus},
	title = "{The properties of the positronium lifetime image reconstruction based on maximum likelihood estimation}",
	url = {http://dx.doi.org/10.5604/01.3001.0054.1807},
	volume = {19},
	author = {Chen, Zhuo and An, Lingling and Kao, Chien-Min and Huang, Hsin-Hsiung},
	pages = {1--8},
	date = {2023-12-31},
	year = {2023},
	month = {Dec.},
	day = {31},
}

@article{FaranakBAMS2024,
	journal = {Bio-Algorithms and Med-Systems},
	doi = {10.5604/01.3001.0054.8095},
	issn = {1896-530X},
	number = {Special Issue},
	publisher = {Index Copernicus},
	title = "{Assessing the Spatial Resolutionof the Modular J-PET Scannerusing the Maximum-LikelihoodExpectation-Maximization (MLEM)algorithm}",
	url = {http://dx.doi.org/10.5604/01.3001.0054.8095},
	volume = {20},
	author = {Tayefi Ardebili, Faranak and Moskal, Paweł},
	pages = {1--9},
	date = {2024-11-21},
	year = {2024},
	month = {Nov.},
	day = {21},
}

@article{Stepanov2020,
   author       = "Stepanov, P. S. and Selim, F. A. and Stepanov, S. V. and Bokov, A. V. and Ilyukhina, O. V. and Duplâtre, G. and Byakov, V. M.",
   title        = "Interaction of positronium with dissolved oxygen in liquids",
   journal      = "Physical Chemistry Chemical Physics",
   year         = "2020",
   volume       = "22",
   pages        = "5123",
   doi          = "10.1039/c9cp06105c"
}

@article{Zgardzinska2020,
   author       = "Zgardzi{\'n}ska, B and Cho{\l}ubek, Gustaw and Jarosz, B and Wysogl{\k{a}}d, Konrad and Gorgol, M and Go{\'z}dziuk, Magdalena and Cho{\l}ubek, Micha{\l} and Jasi{\'n}ska, B",
   title        = "Studies on healthy and neoplastic tissues using positron annihilation lifetime spectroscopy and focused histopathological imaging",
   journal      = "Sci. Rep.",
   year         = "2020",
   volume       = "10",
   pages        = "11890",
   doi          = "10.1038/s41598-020-68727-3"
}

@article{Shopa-Kamil-Bams2023,
   author       = "Shopa, R. and Dulski, K.",
   title        = "Positronium imaging in {J-PET} with an iterative activity reconstruction and a multi-stage fitting algorithm",
   journal      = "Bio-Algorithms and Med-Systems",
   year         = "2023",
   volume       = "19",
   number       = "1",
   pages        = "54-63",
   doi          = "https://doi.org/10.5604/01.3001.0054.1826"
}

@article{Qi-positronium-2022,
   author       = "Qi, J. and Huang, B.",
   title        = "Positronium Lifetime Image Reconstruction for {TOF} {PET}",
   journal      = "IEEE Transactions on Medical Imaging",
   year         = "2022",
   volume       = "41",
   pages        = "2848",
   doi          = "10.1109/TMI.2022.3174561"
}

@ARTICLE{Qi-positronium-2023,
  author={Huang, Bangyan and Li, Tiantian and Ariño-Estrada, Gerard and Dulski, Kamil and Shopa, Roman Y. and Moskal, Pawel and Stępień, Ewa and Qi, Jinyi},
  journal={IEEE Transactions on Medical Imaging}, 
  title="{SPLIT: Statistical Positronium Lifetime Image Reconstruction via Time-Thresholding}", 
  year={2024},
  volume={43},
  number={6},
  pages={2148-2158},
  doi={10.1109/TMI.2024.3357659}}

@article{Steinberger2024,
  author = {Steinberger, W. M. and Mercolli, L. and Breuer, J. and Prenosil, G. and Shi, K. and Rominger, A. and Fischer, U. and Schwyzer, M. and Cumming, P.},
  title = "{Positronium lifetime validation measurements using a long-axial field-of-view positron emission tomography scanner}",
  journal = {EJNMMI Physics},
  volume = {11},
  number = {1},
  pages = {76},
  year = {2024},
  doi = {10.1186/s40658-024-00678-4},
  url = {https://doi.org/10.1186/s40658-024-00678-4},
  issn = {2197-7364},
  publisher = {Springer}
}

@article {Mercolli2024.10.19.24315509,
	author = {Mercolli, Lorenzo and Steinberger, William M. and Sari, Hasan and Afshar-Oromieh, Ali and Caobelli, Federico and Conti, Maurizio and Felgosa Cardoso, {\^A}ngelo R. and Mingels, Clemens and Moskal, Pawe{\l} and Pyka, Thomas and Rathod, Narendra and Schepers, Robin and Seifert, Robert and Shi, Kuangyu and St{\k e}pie{\'n}, Ewa {\L}. and Viscione, Marco and Rominger, Axel O.},
	title = "{In Vivo Positronium Lifetime Measurements with a Long Axial Field-of-View PET/CT}",
	elocation-id = {2024.10.19.24315509},
	year = {2024},
	doi = {10.1101/2024.10.19.24315509},
	publisher = {Cold Spring Harbor Laboratory Press},
	URL = {https://www.medrxiv.org/content/early/2024/10/22/2024.10.19.24315509},
	eprint = {https://www.medrxiv.org/content/early/2024/10/22/2024.10.19.24315509.full.pdf},
	journal = {medRxiv}
}

@INPROCEEDINGS{Huang_mic,
  author={Huang, B. and Dai, B. and Li, E. J. and Karp, J. S. and Qi, J.},
  booktitle={2024 IEEE Nuclear Science Symposium (NSS), Medical Imaging Conference (MIC) and Room Temperature Semiconductor Detector Conference (RTSD)}, 
  title="{High-resolution Positronium Lifetime Imaging on the PennPET Explorer}", 
  year={2024},
  volume={},
  number={},
  pages={1-1},
  doi={10.1109/NSS/MIC/RTSD57108.2024.10656096}}

@article{Duran2022Half,
	journal = {Applied Radiation and Isotopes},
	doi = {10.1016/j.apradiso.2022.110507},
	issn = {0969-8043},
	language = {en},
	publisher = {Elsevier BV},
	title = "{Half-life measurement of 44Sc and 44mSc}",
	url = {http://dx.doi.org/10.1016/j.apradiso.2022.110507},
	volume = {190},
	author = {Durán, M. Teresa and Juget, Frédéric and Nedjadi, Youcef and Bailat, Claude and Grundler, Pascal V. and Talip, Zeynep and van der Meulen, Nicholas P. and Casolaro, Pierluigi and Dellepiane, Gaia and Braccini, Saverio},
	pages = {110507},
	date = {2022-12},
	year = {2022},
	month = {Dec.},
}

@article{Huang2024High,
	journal = {Physics in Medicine \&; Biology},
	doi = {10.1088/1361-6560/ad9543},
	issn = {0031-9155},
	number = {24},
	publisher = {IOP Publishing},
	title = "{High-resolution positronium lifetime tomography by the method of moments}",
	url = {http://dx.doi.org/10.1088/1361-6560/ad9543},
	volume = {69},
	author = {Huang, Bangyan and Qi, Jinyi},
	pages = {24NT01},
	date = {2024-12-04},
	year = {2024},
	month = {Dec.},
	day = {4},
}

@article{ManishBAMS,
   author       = "Das, M. and Mryka, W. and Yitayew, E. and Parzych, S. and Sharma, S. and Stepien, E. and Moskal, P.",
   title        = "Estimating the efficiency and purity for detecting annihilation and prompt photons for positronium imaging with {J-PET} using toy Monte Carlo simulation",
   journal      = "Bio-Algorithms and Med-Systems",
   year         = "2023",
   volume       = "19",
   pages        = "87-95",
   doi          = "10.5604/01.3001.0054.1938"
}

@article{das2024,
  author = {Das, Manish and Bayerlein, Reimund and Sharma, Sushil and Parzych, Szymon and Niedźwiecki, Szymon and Badawi, Ramsey and Beyene, Ermias Yitayew},
  title = "{Development of correction techniques for the J-PET scanner}",
  journal = {Bio-Algorithms and Med-Systems},
  volume = {20},
  number = {1},
  pages = {101--110},
  year = {2024},
  doi = {10.5604/01.3001.0054.9362},
  url = {https://doi.org/10.5604/01.3001.0054.9362}
}

@article{Huang2025Fast,
	journal = {Communications Physics},
	doi = {10.1038/s42005-025-02100-6},
	issn = {2399-3650},
	number = {181},
	language = {en},
	publisher = {Springer Science and Business Media LLC},
	title = "{Fast high-resolution lifetime image reconstruction for positron lifetime tomography}",
	url = {http://dx.doi.org/10.1038/s42005-025-02100-6},
	volume = {8},
	author = {Huang, Bangyan and Wang, Zipai and Zeng, Xinjie and Goldan, Amir H. and Qi, Jinyi},
	date = {2025-04-26},
	year = {2025}
}

@article{Huang2025Statistical,
	journal = {IEEE Transactions on Radiation and Plasma Medical Sciences},
	doi = {10.1109/trpms.2025.3531225},
	issn = {2469-7311},
	number = {4},
	publisher = {Institute of Electrical and Electronics Engineers (IEEE)},
	title = "{A Statistical Reconstruction Algorithm for Positronium Lifetime Imaging Using Time-of-Flight Positron Emission Tomography}",
	url = {http://dx.doi.org/10.1109/trpms.2025.3531225},
	volume = {9},
	author = {Huang, Hsin-Hsiung and Zhu, Zheyuan and Booppasiri, Slun and Chen, Zhuo and Pang, Shuo and Kao, Chien-Min},
	pages = {478--486},
	date = {2025-04},
	year = {2025},
	month = {Apr.},
}

@article{Berens2024analytic,
	journal = {Bio-Algorithms and Med-Systems},
	doi = {10.5604/01.3001.0054.9141},
	issn = {1896-530X},
	number = {Special Issue},
	publisher = {Index Copernicus},
	title = "{An analytic, moment-based method to estimate orthopositronium lifetimes in positron annihilation lifetime spectroscopy measurements}",
	url = {http://dx.doi.org/10.5604/01.3001.0054.9141},
	volume = {20},
	author = {Berens, Lucas and Hsu, Isaac and Chen, Chin-Tu and Halpern, Howard and Kao, Chien-Min},
	pages = {40--48},
	date = {2024-12-23},
	year = {2024},
	month = {Dec.},
	day = {23},
}

@article {Karp136,
	author = {Karp, Joel S. and Viswanath, Varsha and Geagan, Michael J. and Muehllehner, Gerd and Pantel, Austin R. and Parma, Michael J. and Perkins, Amy E. and Schmall, Jeffrey P. and Werner, Matthew E. and Daube-Witherspoon, Margaret E.},
	title = "{PennPET Explorer: Design and Preliminary Performance of a Whole-Body Imager}",
	volume = {61},
	number = {1},
	pages = {136--143},
	year = {2020},
	doi = {10.2967/jnumed.119.229997},
	publisher = {Society of Nuclear Medicine},
	issn = {0161-5505},
	URL = {https://jnm.snmjournals.org/content/61/1/136},
	eprint = {https://jnm.snmjournals.org/content/61/1/136.full.pdf},
	journal = {Journal of Nuclear Medicine}
}

@article{Dai-PennPET-NEMA,
   author       = "Dai, B and Daube-Witherspoon, ME and McDonald, S and Werner, ME and Parma, MJ and Geagan, MJ and Viswanath, V and Karp, JS.",
   title        = "Performance evaluation of the {PennPET} explorer with expanded axial coverage",
   journal      = "Phys. Med. Biol.",
   year         = "2023",
   volume       = "68",
   pages        = "095007 ",
   doi          = "10.1088/1361-6560/acc722"
}

@article{Sitarz2020,
   author       = "Sitarz, M. and Cussonneau, J-P and Matulewicz, T. and Haddad F.",
   title        = "Radionuclide candidates for for $\beta^+$$\gamma$ coincidence {PET}: An overview",
   journal      = "Applied Radiation and Isotopes",
   year         = "2020",
   volume       = "155",
   pages        = "108898",
   doi          = "10.1016/j.apradiso.2019.108898"
}

@article{Moskal-NEMA-PMB-2021,
      author         = "Moskal, P and Kowalski, P and Shopa, RY and Raczyński, L and Baran, J and Chug, N and Curceanu, C and Czerwiński, E and Dadgar, M and Dulski, K and Gajos, A and Hiesmayr, BC and Kacprzak, K and Kapłon, {\L{}} and Kisielewska, D and Klimaszewski, K and Kopka, P and Korcyl, G and Krawczyk, N and Krzemień, W and Kubicz, E and Niedźwiecki, S and Parzych, S and Raj, J and Sharma, S and Shivani, S and Stępień, E and Tayefi, F and Wiślicki, W.",
      title          = "{Simulating NEMA characteristics of the modular total-body J-PET scanner — an economic total-body PET from plastic scintillators}",
      journal        = "Phys. Med. Biol.",
      volume         = "66",
      year           = "2021",
      pages          = "175015",
      doi            = "10.1088/1361-6560/ac16bd"
      }

@article{Jean2006,
  author =      "Jean, Y.C. and Li, Y. and Liu, G. and Chen, H. and Zhang J. and Gadzia, J.E.",
  title =       "Applications of slow positrons to cancer research: Search for selectivity of positron annihilation to skin cancer",
  journal =     "Applied Surface Science",
  volume  =     "252",
  pages   =     "3166",
  year    =     "2006",
  doi     =     "https://doi.org/10.1016/j.apsusc.2005.08.101"
}

@article{Chen2012,
  author =      "Chen, H. and van Horn, J.D. and Jean, Y.C.",
  title =       "Applications of positron annihilation spectroscopy to life science",
  journal =     "Defect Diffusion Forum",
  volume  =     "331",
  pages   =     "275",
  year    =     "2012",
  doi     =     "https://doi.org/10.4028/www.scientific.net/DDF.331.275"
}

@article{Jasinska2017,
  author =      "Jasińska, B. and Zgardzińska, B. and Chołubek, G. and Gorgol, M. and Wiktor, K. and Wysogląd, K. and Białas, P. and Curceanu, C. and Czerwiński, E. and Dulski, K. and Gajos, A. and Głowacz, B. and Hiesmayr, B. and Jodłowska-Jędrych, B. and Kamińska, D. and Korcyl, G. and Kowalski, P. and Kozik, T. and Krawczyk, N. and Krzemień, W. and Kubicz, E. and Mohammed, M. and Pawlik-Niedźwiecka, M. and Niedźwiecki, S. and Pałka, M. and Raczyński, L. and Rudy, Z. and Sharma, N.G. and Sharma, S. and Shopa, R. and Silarski, M. and Skurzok, M. and Wieczorek, A. and Wiktor, H. and Wiślicki, W. and Zieliński, M. and Moskal, P.",
  title =       "Human tissues investigation using PALS technique",
  journal =     "Acta Phys. Polon. B",
  volume  =     "48",
  pages   =     "1737",
  year    =     "2017",
  doi     =     "https://doi.org/10.5506/APhysPolB.48.1737"
}

@article{EJNMMI2023,
  title={Developing a novel positronium biomarker for cardiac myxoma imaging},
  author={Moskal, Pawe{\l} and Kubicz, Ewelina and Grudzie{\'n}, Grzegorz and Czerwi{\'n}ski, Eryk and Dulski, Kamil and Leszczy{\'n}ski, Bartosz and Nied{\'z}wiecki, Szymon and St{\k{e}}pie{\'n}, Ewa {\L}},
  journal={EJNMMI Physics},
  volume={10},
  number={1},
  pages={22},
  year={2023},
publisher={Springer},
  doi = {10.1186/s40658-023-00543-w}
  
}

@article{Ahn-Gidley-colagen-2021,
  author =      "Ahn, Taeyong and Gidley, David W. and Thornton, Aaron W. and Wong-Foy, Antek G. and Orr, Bradford G. and Kozloff, Kenneth M. and Banaszak Holl, Mark M.",
  title =       "Hierarchical Nature of Nanoscale Porosity in Bone Revealed by Positron Annihilation Lifetime Spectroscopy",
  journal =     "ACS Nano",
  volume  =     "15",
  pages   =     "4321",
  year    =     "2021",
  doi     =     "10.1021/acsnano.0c07478"
}

@article{Karimi2023,
  author =      "Karimi, H. and Moskal, P. and Zak, A. and Stepien, E.",
  title =       "3D melanoma spheroid model for the development of positronium biomarker",
  journal =     "Sci. Rep.",
  volume  =     "13",
  pages   =     "1648",
  year    =     "2023",
  doi     =     "10.1038/s41598-023-34571-4"
}

@inPROCEEDINGS{IEEEpositronium-imaging,
  author={Moskal, Pawel},
  booktitle={2019 IEEE Nuclear Science Symposium and Medical Imaging Conference (NSS/MIC)}, 
  title={Positronium Imaging}, 
  year={2019},
  volume={},
  number={},
  pages={1-3},
  doi={10.1109/NSS/MIC42101.2019.9059856}}

@article{BAMShypoxia2021,
  author =      "Moskal, P. and Stepien, E.",
  title =       "Positronium as a biomarker of hypoxia",
  journal =     "Bio-Algorithms and Med-Systems",
  volume  =     "17",
  pages   =     "311",
  year    =     "2021",
  doi     =     "10.1515/bams-2021-0189"
}

@article{TAKYU2024169514,
title = "{Positronium lifetime measurement using a clinical PET system for tumor hypoxia identification}",
journal = {Nuclear Instruments and Methods in Physics Research Section A: Accelerators, Spectrometers, Detectors and Associated Equipment},
volume = {1065},
pages = {169514},
year = {2024},
issn = {0168-9002},
doi = {https://doi.org/10.1016/j.nima.2024.169514},
url = {https://www.sciencedirect.com/science/article/pii/S0168900224004406},
author = {Sodai Takyu and Fumihiko Nishikido and Hideaki Tashima and Go Akamatsu and Ken-ichiro Matsumoto and Miwako Takahashi and Taiga Yamaya}
}

@article{Takyu_2024,
doi = {10.35848/1347-4065/ad679a},
url = {https://dx.doi.org/10.35848/1347-4065/ad679a},
year = {2024},
month = {Aug.},
publisher = {IOP Publishing},
volume = {63},
number = {8},
pages = {086003},
author = {Takyu, Sodai and Matsumoto, Ken-ichiro and Hirade, Tetsuya and Nishikido, Fumihiko and Akamatsu, Go and Tashima, Hideaki and Takahashi, Miwako and Yamaya, Taiga},
title = "{Quantification of radicals in aqueous solution by positronium lifetime: an experiment using a clinical PET scanner}",
journal = {Japanese Journal of Applied Physics}
}

@article {Huangjnumed.125.270130,
  title = "{High-Resolution Positronium Lifetime Tomography at Clinical Activity Levels on the PennPET Explorer}",
  volume = {66},
  ISSN = {2159-662X},
  url = {http://dx.doi.org/10.2967/jnumed.125.270130},
  DOI = {10.2967/jnumed.125.270130},
  number = {9},
  journal = {Journal of Nuclear Medicine},
  publisher = {Society of Nuclear Medicine},
  author = {Huang,  Bangyan and Dai,  Bing and Lapi,  Suzanne E. and Liles,  Grace and Karp,  Joel S. and Qi,  Jinyi},
  year = {2025},
  month = aug,
  pages = {1464–1470}
}

@article{Shibuya2020,
  author =      "Shibuya, K. and Saito, H. and Nishikido, F. and Takahashi, M. and Yamaya, T.",
  title =       "Oxygen sensing ability of positronium atom for tumor hypoxia imaging",
  journal =     "Commun. Phys.",
  volume  =     "3",
  pages   =     "173",
  year    =     "2020",
  doi     =     "10.1038/s42005-020-00440-z"
}

@article{RMP2023,
  author =      "Bass, S. and Mariazzi, S. and Moskal, P. and Stepien, E.",
  title =       "Positronium physics and biomedical applications",
  journal =     "Rev. Mod. Phys.",
  volume  =     "95",
  pages   =     "021002",
  year    =     "2023",
  doi     =     "10.1103/RevModPhys.95.021002"
}

@article{Spencer-uExplorer-NEMA-2021,
      author         = "Benjamin A Spencer and Eric Berg and Jeffrey P Schmall and Negar Omidvari and Edwin K Leung and Yasser G Abdelhafez and Songsong Tang and Zilin Deng and Yun Dong and Yang Lv and Jun Bao and Weiping Liu and Hongdi Li and Terry Jones and Ramsey D Badawi and Simon R Cherry",
      title          = "{Performance evaluation of the uEXPLORER total-body PET/CT scanner based on NEMA NU 2-2018 with additional tests to characterize PET scanners with a long axial field of view}",
      journal        = "J. Nucl. Med.",
      volume         = "61",
      year           = "2021",
      pages          = "861",
      doi            = "10.2967/jnumed.120.250597"
      }

@article{Prenosil-quadra-2022,
      author         = "G A Prenosil and H Sari and M F{\"u}rstner and A Afshar-Oromieh and K Shi and A Rominger and M Hentschel",
      title          = "{Performance characteristics of the biograph vision Quadra {PET/CT} system with a long axial field of view using the NEMA NU 2-2018 standard}",
      journal        = "J. Nucl. Med.",
      volume         = "63",
      year           = "2022",
      pages          = "476",
      doi            = "10.2967/jnumed.121.261972"
      }

@article{Alavi:2021,
  author =       "Alavi, Abass and Werner, Thomas J. and St\k{e}pie{\'n}, Ewa {\L{}}. and Moskal, Pawel",
  title =        "{Unparalleled and revolutionary impact of PET imaging on research and day to day practice of medicine}",
  journal =      "Bio-Algorithms and Med-Systems",
  year =         "2021",
  volume =       "17",
  pages =        "203-220",
doi            = "https://doi.org/10.1515/bams-2021-0186",
}

@article{Sharma:2020,
  author =       "S. Sharma and J. Chhokar and C. Curceanu and E. Czerwinski and M. Dadgar and K. Dulski and J. Gajewski and A. Gajos and M. Gorgol and N. Gupta-Sharma and R. Del Grande and B. C. Hiesmayr and B. Jasinska and K. Kacprzak and {\L}. Kaplon and H. Karimi and D. Kisielewska and K. Klimaszewski and G. Korcyl and P. Kowalski and T. Kozik and N. Krawczyk and W. Krzemien and E. Kubicz and M. Mohammed and Sz. Niedzwiecki and M. Palka and M. Pawlik-Niedzwiecka and L. Raczynski and J. Raj and A. Rucinski and Shivani and R. Y. Shopa and M. Silarski and M. Skurzok and E. {\L}. Stepie{\'n} and W. Wislicki and B. Zgardzinska and P. Moskal",
  title =        "{Estimating relationship between the Time Over
                  Threshold and energy loss by photons in plastic
                  scintillators used in the J-PET scanner}",
  journal =      "EJNMMI Phys.",
  year =         "2020",
  volume =       "7",
  pages =        "39",
  archivePrefix ="arXiv",
  primaryClass = "physics.ins-det",
  SLACcitation = "%%CITATION = ARXIV:1911.12059;%%",
  doi = "https://doi.org/10.1186/s40658-020-00306-x"
}

@article{NIM2014,
      author         = "P. Moskal and Sz. Niedźwiecki and T. Bednarski and E. Czerwiński and {\L{}} Kapłon and E. Kubicz and I. Moskal and M. Pawlik-Niedźwiecka and N.G. Sharma and M. Silarski and M. Zieliński and N. Zoń and P. Białas and A. Gajos and A. Kochanowski and G. Korcyl and J. Kowal and P. Kowalski and T. Kozik and W. Krzemień and M. Molenda and M. Pałka and L. Raczyński and Z. Rudy and P. Salabura and A. Słomski and J. Smyrski and A. Strzelecki and A. Wieczorek and W. Wiślicki",
      title          = "Test of a single module of the {J-PET} scanner based on plastic scintillators",
      journal        = "Nucl. Instr. and Meth. A",
      volume         = "764",
      year           = "2014",
      pages          = "317-321",
      doi            = "10.1016/j.nima.2014.07.052",
      eprint         = "1407.7395",
      archivePrefix  = "arXiv",
      primaryClass   = "physics.ins-det",
      SLACcitation   = "%%CITATION = ARXIV:1407.7395;%%"
}

@article{PMB2019,
      author         = "P Moskal and D Kisielewska and C Curceanu and E Czerwiński and K Dulski and A Gajos and M Gorgol and B Hiesmayr and B Jasińska and K Kacprzak and {\L{}} Kapłon and G Korcyl and P Kowalski and W Krzemień and T Kozik and E Kubicz and M Mohammed and Sz Niedźwiecki and M Pałka and M Pawlik-Niedźwiecka and L Raczyński and J Raj and S Sharma and  Shivani and R Y Shopa and M Silarski and M Skurzok and E Stępień and W Wiślicki and B Zgardzińska",
      title          = "{Feasibility study of the positronium imaging with the
                        {J-PET} tomograph}",
      journal        = "Phys. Med. Biol.",
      volume         = "64",
      year           = "2019",
      number         = "5",
      pages          = "055017",
      doi            = "10.1088/1361-6560/aafe20"
}

@article{Niedzwiecki:2017nka,
      author         = "S. Niedźwiecki and P. Białas and C. Curceanu and E. Czerwiński and K. Dulski and A. Gajos and B. Głowacz and M. Gorgol and B. C. Hiesmayr and B. Jasińska and {\L{}} Kapłon and D. Kisielewska-Kamińska and G. Korcyl and P. Kowalski and T. Kozik and N. Krawczyk and W. Krzemień and E. Kubicz and M. Mohammed and M. Pawlik-Niedźwiecka and M. Pałka and L. Raczyński and Z. Rudy and N. G. Sharma and S. Sharma and R. Y. Shopa and M. Silarski and M. Skurzok and A. Wieczorek and W. Wiślicki and B. Zgardzińska and M. Zieliński and P. Moskal",
      title          = "{J-PET: a new technology for the whole-body PET imaging}",
      journal        = "Acta Phys. Polon. B",
      volume         = "48",
      year           = "2017",
      pages          = "1567",
      doi            = "10.5506/APhysPolB.48.1567"
}

@inproceedings{Samanta2023Feasibility,
	author = {Samanta, S. and Sun, X. and Li, H. and Li, Y.},
	booktitle = {2023 {IEEE} {Nuclear} {Science} {Symposium}, {Medical} {Imaging} {Conference} and {International} {Symposium} on {Room}-{Temperature} {Semiconductor} {Detectors} ({NSS} {MIC} {RTSD})},
	doi = {10.1109/nssmicrtsd49126.2023.10338538},
	year = {2023},
	month = {nov 4},
	pages = {1--1},
	organization = {IEEE},
	title = "{Feasibility {Study} of {Positronium} {Imaging} using the {NeuroEXPLORER} ({NX}) {Brain} {PET} {Scanner}}",
	url = {http://dx.doi.org/10.1109/NSSMICRTSD49126.2023.10338538},
}

@ARTICLE{korcyl_ieee,
  author =	 {Korcyl, G. and Białas, P. and Curceanu, C. and Czerwiński, E. and Dulski, K. and Flak, B. and Gajos, A. and Głowacz, B. and Gorgol, M. and Hiesmayr, B. C. and Jasińska, B. and Kacprzak, K. and Kajetanowicz, M. and Kisielewska, D. and Kowalski, P. and Kozik, T. and Krawczyk, N. and Krzemień, W. and Kubicz, E. and Mohammed, M. and Niedźwiecki, Sz. and Pawlik-Niedźwiecka, M. and Pałka, M. and Raczyński, L. and Rajda, P. and Rudy, Z. and Salabura, P. and Sharma, N. G. and Sharma, S. and Shopa, R. Y. and Skurzok, M. and Silarski, M. and Strzempek, P. and Wieczorek, A. and Wiślicki, W. and Zaleski, R. and Zgardzińska, B. and Zieliński, M. and Moskal, P.},
  journal =	 {IEEE Transactions on Medical Imaging},
  title =	 {Evaluation of Single-Chip, Real-Time Tomographic
                  Data Processing on FPGA SoC Devices},
  year =	 {2018},
  volume =	 {37},
  number =	 {11},
  pages =	 {2526-2535},
  doi =		 "10.1109/TMI.2018.2837741"
}

@article{EJNMMI2020,
  author =       "Moskal, P. and Kisielewska, D. and Shopa, R. Y. and Bura, Z. and
   Chhokar, J. and Curceanu, C. and Czerwinski, E. and Dadgar, M. and
   Dulski, K. and Gajewski, J. and Gajos, A. and Gorgol, M. and Del Grande,
   R. and Hiesmayr, B. C. and Jasinska, B. and Kacprzak, K. and Kaminska,
   A. and Kaplon, L. and Karimi, H. and Korcyl, G. and Kowalski, P. and
   Krawczyk, N. and Krzemien, W. and Kozik, T. and Kubicz, E. and Malczak,
   P. and Mohammed, M. and Niedzwiecki, Sz and Palka, M. and
   Pawlik-Niedzwiecka, M. and Pedziwiatr, M. and Raczynski, L. and Raj, J.
   and Rucinski, A. and Sharma, S. and Shivani, S. and Silarski, M. and
   Skurzok, M. and Stepien, E. L. and Vandenberghe, S. and Wislicki, W. and
   Zgardzinska, B.",
  title =        "{Performance assessment of the 2$\gamma$ positronium
                  imaging with the total-body PET scanners}",
  year =         "2020",
  journal =      "EJNMMI Phys.",
  volume =       "7",
  pages =        "44",
  url = {http://dx.doi.org/10.1186/s40658-020-00307-w},
  DOI = {10.1186/s40658-020-00307-w}
}

@article{Moskal2024Discrete,
	journal = {Nature Communications},
	doi = {10.1038/s41467-023-44340-6},
	issn = {2041-1723},
	number = {78},
	language = {en},
	publisher = {Springer Science and Business Media LLC},
	title = "{Discrete symmetries tested at 10$-$4 precision using linear polarization of photons from positronium annihilations}",
	url = {http://dx.doi.org/10.1038/s41467-023-44340-6},
	volume = {15},
	author = {Moskal, Paweł and Czerwiński, Eryk and Raj, Juhi and Bass, Steven D. and Beyene, Ermias Y. and Chug, Neha and Coussat, Aurélien and Curceanu, Catalina and Dadgar, Meysam and Das, Manish and Dulski, Kamil and Gajos, Aleksander and Gorgol, Marek and Hiesmayr, Beatrix C. and Jasińska, Bożena and Kacprzak, Krzysztof and Kaplanoglu, Tevfik and Kapłon, {\L{}}ukasz and Klimaszewski, Konrad and Konieczka, Paweł and Korcyl, Grzegorz and Kozik, Tomasz and Krzemień, Wojciech and Kumar, Deepak and Moyo, Simbarashe and Mryka, Wiktor and Niedźwiecki, Szymon and Parzych, Szymon and del Río, Elena Pérez and Raczyński, Lech and Sharma, Sushil and Choudhary, Shivani and Shopa, Roman Y. and Silarski, Michał and Skurzok, Magdalena and Stępień, Ewa {\L{}} and Tanty, Pooja and Ardebili, Faranak Tayefi and Ardebili, Keyvan Tayefi and Eliyan, Kavya Valsan and Wiślicki, Wojciech},
	date = {2024-01-02},
	year = {2024}
}

@article{pet-clinics,
  author =       "Moskal, P. and St\k{e}pie{\'n}, E. {\L{}}.",
  journal =      "PET Clinics",
  volume =       "15",
  year =         "2020",
  pages =        "439--452",
  title =        "{Prospects and clinical perspectives of total-body
                  PET imaging using plastic scintillators}",
  doi = " 10.1016/j.cpet.2020.06.009"
}

@article{ doi:10.1126/sciadv.abh4394,
  author =       "Paweł Moskal  and Kamil Dulski  and Neha Chug  and Catalina Curceanu  and Eryk Czerwiński  and Meysam Dadgar  and Jan Gajewski  and Aleksander Gajos  and Grzegorz Grudzień  and Beatrix C. Hiesmayr  and Krzysztof Kacprzak  and {\L{}}ukasz Kapłon  and Hanieh Karimi  and Konrad Klimaszewski  and Grzegorz Korcyl  and Paweł Kowalski  and Tomasz Kozik  and Nikodem Krawczyk  and Wojciech Krzemień  and Ewelina Kubicz  and Piotr Małczak  and Szymon Niedźwiecki  and Monika Pawlik-Niedźwiecka  and Michał Pędziwiatr  and Lech Raczyński  and Juhi Raj  and Antoni Ruciński  and Sushil Sharma  and Shivani null  and Roman Y. Shopa  and Michał Silarski  and Magdalena Skurzok  and Ewa {\L{}} Stępień  and Monika Szczepanek  and Faranak Tayefi  and Wojciech Wiślicki",
  title =        "Positronium imaging with the novel multiphoton {PET} scanner",
  journal =      "Science Advances",
  volume =       "7",
  number =       "42",
  pages =        "eabh4394",
  year =         "2021",
  doi =          "10.1126/sciadv.abh4394"
}

@article{Moyo2022Feasibility,
	journal = {Bio-Algorithms and Med-Systems},
	doi = {10.2478/bioal-2022-0087},
	issn = {1896-530X},
	number = {1},
	language = {en},
	publisher = {Index Copernicus},
	title = "{Feasibility study of positronium application for blood clots structural characteristics}",
	url = {http://dx.doi.org/10.2478/bioal-2022-0087},
	volume = {18},
	author = {Moyo, Simbarashe and Moskal, Paweł and Stępień, Ewa {\L{}}},
	pages = {163--167},
	date = {2022-12-01},
	year = {2022},
	month = {12},
	day = {1},
}

@article{Moskal2025Nonmaximal,
author = {Paweł Moskal  and Deepak Kumar  and Sushil Sharma  and Ermias Yitayew Beyene  and Neha Chug  and Aurélien Coussat  and Catalina Curceanu  and Eryk Czerwiński  and Manish Das  and Kamil Dulski  and Marek Gorgol  and Bożena Jasińska  and Krzysztof Kacprzak  and Tevfik Kaplanoglu  and Łukasz Kapłon  and Tomasz Kozik  and Edward Lisowski  and Filip Lisowski  and Wiktor Mryka  and Szymon Niedźwiecki  and Szymon Parzych  and Elena P. del Rio  and Martin Rädler  and Magdalena Skurzok  and Ewa {\L{}} Stepień  and Pooja Tanty  and Keyvan Tayefi Ardebili  and Kavya Valsan Eliyan },
title = "{Nonmaximal entanglement of photons from positron-electron annihilation demonstrated using a plastic PET scanner}",
journal = {Science Advances},
volume = {11},
number = {18},
pages = {eads3046},
year = {2025},
doi = {10.1126/sciadv.ads3046},
URL = {https://www.science.org/doi/abs/10.1126/sciadv.ads3046},
eprint = {https://www.science.org/doi/pdf/10.1126/sciadv.ads3046}}

@article{
doi:10.1126/sciadv.adp2840,
author = {Paweł Moskal  and Jakub Baran  and Steven Bass  and Jarosław Choiński  and Neha Chug  and Catalina Curceanu  and Eryk Czerwiński  and Meysam Dadgar  and Manish Das  and Kamil Dulski  and Kavya V. Eliyan  and Katarzyna Fronczewska  and Aleksander Gajos  and Krzysztof Kacprzak  and Marcin Kajetanowicz  and Tevfik Kaplanoglu  and {\L{}}ukasz Kapłon  and Konrad Klimaszewski  and Małgorzata Kobylecka  and Grzegorz Korcyl  and Tomasz Kozik  and Wojciech Krzemień  and Karol Kubat  and Deepak Kumar  and Jolanta Kunikowska  and Joanna Mączewska  and Wojciech Migdał  and Gabriel Moskal  and Wiktor Mryka  and Szymon Niedźwiecki  and Szymon Parzych  and Elena P. del Rio  and Lech Raczyński  and Sushil Sharma  and Shivani Shivani  and Roman Y. Shopa  and Michał Silarski  and Magdalena Skurzok  and Faranak Tayefi  and Keyvan T. Ardebili  and Pooja Tanty  and Wojciech Wiślicki  and Leszek Królicki  and Ewa {\L{}} Stępień },
title = "{Positronium image of the human brain in vivo}",
journal = {Science Advances},
volume = {10},
number = {37},
pages = {eadp2840},
year = {2024},
doi = {10.1126/sciadv.adp2840},
URL = {https://www.science.org/doi/abs/10.1126/sciadv.adp2840},
eprint = {https://www.science.org/doi/pdf/10.1126/sciadv.adp2840}}

@article{Moskal:2021kxe,
	journal = {Nature Communications},
	doi = {10.1038/s41467-021-25905-9},
	issn = {2041-1723},
	number = {5658},
	language = {en},
	publisher = {Springer Science and Business Media LLC},
	title = "{Testing CPT symmetry in ortho-positronium decays with positronium annihilation tomography}",
	url = {http://dx.doi.org/10.1038/s41467-021-25905-9},
	volume = {12},
	author = {Moskal, P. and Gajos, A. and Mohammed, M. and Chhokar, J. and Chug, N. and Curceanu, C. and Czerwiński, E. and Dadgar, M. and Dulski, K. and Gorgol, M. and Goworek, J. and Hiesmayr, B. C. and Jasińska, B. and Kacprzak, K. and Kapłon, {\L{}} and Karimi, H. and Kisielewska, D. and Klimaszewski, K. and Korcyl, G. and Kowalski, P. and Krawczyk, N. and Krzemień, W. and Kozik, T. and Kubicz, E. and Niedźwiecki, S. and Parzych, S. and Pawlik-Niedźwiecka, M. and Raczyński, L. and Raj, J. and Sharma, S. and Choudhary, S. and Shopa, R. Y. and Sienkiewicz, A. and Silarski, M. and Skurzok, M. and Stępień, E. {\L{}} and Tayefi, F. and Wiślicki, W.},
	date = {2021-09-27},
	year = {2021}
}

@article{Moskal2025IEEE,
  title = {Positronium Imaging: History,  Current Status,  and Future Perspectives},
  volume = {9},
  ISSN = {2469-7303},
  url = {http://dx.doi.org/10.1109/trpms.2025.3583554},
  DOI = {10.1109/trpms.2025.3583554},
  number = {8},
  journal = {IEEE Transactions on Radiation and Plasma Medical Sciences},
  publisher = {Institute of Electrical and Electronics Engineers (IEEE)},
  author = {Moskal,  Paweł and Bilewicz,  Aleksander and Das,  Manish and Huang,  Bangyan and Khreptak,  Aleksander and Parzych,  Szymon and Qi,  Jinyi and Rominger,  Axel and Seifert,  Robert and Sharma,  Sushil and Shi,  Kuangyu and Steinberger,  William M. and Walczak,  Rafał and Stępień,  Ewa},
  year = {2025},
  month = nov,
  pages = {981–1001}
}

@article{Moskal2024vision,
	journal = {Bio-Algorithms and Med-Systems},
	doi = {10.5604/01.3001.0054.9273},
	issn = {1896-530X},
	number = {Special Issue},
	publisher = {Index Copernicus},
	title = "{A vision to increase the availability of PET diagnostics in low- and medium-income countries by combining a low-cost modular J-PET tomograph with the 44Ti/44Sc generator}",
	url = {http://dx.doi.org/10.5604/01.3001.0054.9273},
	volume = {20},
	author = {Moskal, Paweł and Stępień, Ewa and Khreptak, Aleksander},
	pages = {55--62},
	date = {2024-12-27},
	year = {2024},
	month = {Dec.},
	day = {27},
}

@article{PALSAva1,
  title={Analysis procedure of the positronium lifetime spectra for the {J-PET} detector},
  author={K. Dulski and B. Zgardzińska and P. Bia{\l{}}as and C. Curceanu and E. Czerwiński and A. Gajos and B. G{\l{}}owacz and M. Gorgol and B. C. Hiesmayr and B. Jasińska and D. Kisielewska-Kamińska and G. Korcyl and P. Kowalski and T. Kozik and N. Krawczyk and W. Krzemień and E. Kubicz and M. Mohammed and M. Pawlik-Niedźwieckaand S. Niedźwiecki and M. Pa{\l{}}ka and L. Raczyński and J. Raj and Z. Rudy and N. G. Sharma and S. Sharma and Shivani and R. Y. Shopa and M. Silarski and M. Skurzok and A. Wieczorek and W. Wiślicki and M. Zieliński and P. Moskal},
  journal={Acta Phys. Polon. A},
  volume={132},
  doi={10.12693/APhysPolA.132.1637},
  number={5},
  pages={1637-1640},
  year={2017},
  publisher={Via Medica}
}

@article{PALSAva2,
  title={Commissioning of the {J-PET} detector in view of the positron annihilation lifetime spectroscopy},
  author={K. Dulski and C. Curceanu and E. Czerwiński and A. Gajos and M. Gorgol and N. Gupta-Sharma and B. C. Hiesmayr and B. Jasińska and K. Kacprzak and {\L{}}. Kap{\l{}}on and D. Kisielewska and K. Klimaszewski and G. Korcyl and P. Kowalski and N. Krawczyk and W. Krzemień and T. Kozik and E. Kubicz and M. Mohammed and Sz. Niedźwiecki and M. Pa{\l{}}ka and M. Pawlik-Niedźwiecka and L. Raczyński and J. Raj and K. Rakoczy and Z. Rudy and S. Sharma and Shivani and R. Y. Shopa and M. Silarski and M. Skurzok and W. Wiślicki and B. Zgardzińska and P. Moskal},
  journal={Hyperfine Interactions},
  volume={239},
  doi={10.1007/s10751-018-1517-z},
  number={1},
  pages={40},
  year={2020},
  publisher={Springer}
}

@article{PALSAva3,
  title={{PALS} Avalanche - A New {PAL} Spectra Analysis Software},
  author={K. Dulski},
  journal={Acta Phys. Polon. A},
  volume={137},
  doi={10.12693/APhysPolA.137.167},
  number={2},
  pages={167-170},
  year={2020}
}

@article {Phelps661,
	author = {Phelps, Michael E.},
	title = "{PET: The Merging of Biology and Imaging into Molecular Imaging}",
	volume = {41},
	number = {4},
	pages = {661--681},
	year = {2000},
	publisher = {Society of Nuclear Medicine},
	issn = {0161-5505},
	URL = {https://jnm.snmjournals.org/content/41/4/661},
	eprint = {https://jnm.snmjournals.org/content/41/4/661.full.pdf},
	journal = {Journal of Nuclear Medicine}
}

@book{jean:2003principles,
  editor    = {Y.~C. Jean and P.~E. Mallon and D.~M. Schrader},
  title     = {Principles and Applications of Positron and Positronium Chemistry},
  publisher = {World Scientific},
  year      = {2003},
  isbn      = {978-981-277-561-0},
  series    = {Advanced Series in Physical Chemistry},
  volume    = {12},
  doi       = {10.1142/5086}
}

@article{Gidley2006,
   author = "Gidley, David W. and Peng, Hua-Gen and Vallery, Richard S.",
   title = "POSITRON ANNIHILATION AS A METHOD TO CHARACTERIZE POROUS MATERIALS", 
   journal= "Annual Review of Materials Research",
   year = "2006",
   volume = "36",
   number = "Volume 36, 2006",
   pages = "49-79",
   doi = "https://doi.org/10.1146/annurev.matsci.36.111904.135144",
   url = "https://www.annualreviews.org/content/journals/10.1146/annurev.matsci.36.111904.135144",
   publisher = "Annual Reviews",
   issn = "1545-4118",
   type = "Journal Article",
  }

@article{Mercolli2025,
  title = "{Phantom imaging demonstration of positronium lifetime with a long axial field-of-view PET/CT and 124I}",
  volume = {12},
  ISSN = {2197-7364},
  url = {http://dx.doi.org/10.1186/s40658-025-00790-z},
  DOI = {10.1186/s40658-025-00790-z},
  number = {1},
  journal = {EJNMMI Physics},
  publisher = {Springer Science and Business Media LLC},
  author = {Mercolli,  Lorenzo and Steinberger,  William M. and Rathod,  Narendra and Conti,  Maurizio and Moskal,  Paweł and Rominger,  Axel and Seifert,  Robert and Shi,  Kuangyu and Stępień,  Ewa Ł. and Sari,  Hasan},
  year = {2025},
  month = aug 
}

@article{Mercolli2025IODINE,
  title = "{In vivo voxel-wise positronium lifetime imaging of thyroid cancer using clinically routine I-124 PET/CT}",
  ISSN = {3051-2913},
  url = {http://dx.doi.org/10.1016/j.eanmi.2025.100017},
  DOI = {10.1016/j.eanmi.2025.100017},
  journal = {EANM Innovation},
  publisher = {Elsevier BV},
  author = {Mercolli,  Lorenzo and Steinberger,  William M. and L\"{a}ppchen,  Tilman and Amon,  Michelle and Bregenzer,  Carola and Conti,  Maurizio and Cardoso,  undefinedngelo R. Felgosa and Mingels,  Clemens and Moskal,  Paweł and Rathod,  Narendra and Sari,  Hasan and Stępień,  Ewa Ł. and Weidner,  Sabine and Rominger,  Axel and Shi,  Kuangyu and Seifert,  Robert},
  year = {2025},
  month = dec,
  pages = {100017}
}

@article{1Co2017,
  title = "{A PSMA Ligand Labeled with Cobalt-55 for PET Imaging of Prostate Cancer}",
  volume = {19},
  ISSN = {1860-2002},
  url = {http://dx.doi.org/10.1007/s11307-017-1121-7},
  DOI = {10.1007/s11307-017-1121-7},
  number = {6},
  journal = {Molecular Imaging and Biology},
  publisher = {Springer Science and Business Media LLC},
  author = {Dam,  Johan Hygum and Olsen,  Birgitte Brinkmann and Baun,  Christina and Høilund-Carlsen,  Poul Flemming and Thisgaard,  Helge},
  year = {2017},
  month = sep,
  pages = {915–922}
}

@article{Cobalt_production,
  title = "{Production,  Purification,  and Applications of a Potential Theranostic Pair: Cobalt-55 and Cobalt-58m}",
  volume = {11},
  ISSN = {2075-4418},
  url = {http://dx.doi.org/10.3390/diagnostics11071235},
  DOI = {10.3390/diagnostics11071235},
  number = {7},
  journal = {Diagnostics},
  publisher = {MDPI AG},
  author = {Barrett,  Kendall E. and Houson,  Hailey A. and Lin,  Wilson and Lapi,  Suzanne E. and Engle,  Jonathan W.},
  year = {2021},
  month = jul,
  pages = {1235}
}

@misc{Co_multi,
  doi = {10.48550/ARXIV.2411.08237},
  url = {https://arxiv.org/abs/2411.08237},
  author = {Zou,  Sarah J and Lim,  Irene and Foster,  Jackson W and Chinn,  Garry and Houson,  Hailey A and Lapi,  Suzanne E. and Rao,  Jianghong and Levin,  Craig S},
  title = {Quantitative Imaging of $^{55}\text{Co}$ and $^{18}\text{F}$-Labeled Tracers in a Single "Multiplexed" PET Imaging Session},
  publisher = {arXiv},
  year = {2024},
  copyright = {Creative Commons Attribution 4.0 International}
}

@inproceedings{Co_multi2,
  title = "{Quantitative Imaging of 55CO and 18F-Labeled Tracers in a Single “Multiplexed” Pet Imaging Session}",
  url = {http://dx.doi.org/10.1109/ISBI60581.2025.10980731},
  DOI = {10.1109/isbi60581.2025.10980731},
  booktitle = {2025 IEEE 22nd International Symposium on Biomedical Imaging (ISBI)},
  publisher = {IEEE},
  author = {Zou,  Sarah J. and Lim,  Irene and Foster,  Jackson W. and Chinn,  Garry and Houson,  Hailey A. and Lapi,  Suzanne E. and Rao,  Jianghong and Levin,  Craig S.},
  year = {2025},
  month = apr,
  pages = {1–5}
}

@article{Co_braad,
doi = {10.1088/0031-9155/60/9/3479},
url = {https://dx.doi.org/10.1088/0031-9155/60/9/3479},
year = {2015},
month = {apr},
publisher = {IOP Publishing},
volume = {60},
number = {9},
pages = {3479},
author = {Braad, P E N and Hansen, S B and Thisgaard, H and Høilund-Carlsen, P F},
title = {PET imaging with the non-pure positron emitters: 55Co, 86Y and 124I},
journal = {Physics in Medicine \& Biology}
}

@article{Houson2022Co,
	author = {Houson, Hailey A. and Tekin, Volkan and Lin, Wilson and Aluicio-Sarduy, Eduardo and Engle, Jonathan W. and Lapi, Suzanne E.},
	journal = {Pharmaceutics},
	doi = {10.3390/pharmaceutics14122724},
	issn = {1999-4923},
	number = {12},
	year = {2022},
	month = {dec 6},
	pages = {2724},
	publisher = {MDPI AG},
	title = {PET {Imaging} of the {Neurotensin} {Targeting} {Peptide} {NOTA}-{NT}-20.3 {Using} {Cobalt}-55, {Copper}-64 and {Gallium}-68},
	url = {http://dx.doi.org/10.3390/pharmaceutics14122724},
	volume = {14},
}

@article{Stevens199955Cobalt,
	author = {Stevens, Henk and Jansen, Hugo M.L and De Reuck, Jacques and Lemmerling, Marc and Strijckmans, Karel and Goethals, Patrick and Lemahieu, Ignace and de Jong, Bauke M and Willemsen, Antoon T.M and Korf, Jakob},
	journal = {Journal of the Neurological Sciences},
	doi = {10.1016/s0022-510x(99)00229-4},
	issn = {0022-510X},
	number = {1},
	year = {1999},
	month = {12},
	pages = {11--18},
	publisher = {Elsevier BV},
	title = {55Cobalt ({Co}) as a {PET}-tracer in stroke, compared with blood flow, oxygen metabolism, blood volume and gadolinium-{MRI}},
	url = {http://dx.doi.org/10.1016/S0022-510X(99)00229-4},
	volume = {171},
}

@article{Wooten2015Mnproduction,
  title = "{Cross-sections for (p, x) reactions on natural chromium for the production of 52, 52m, 54Mn radioisotopes}",
  volume = {96},
  ISSN = {0969-8043},
  url = {http://dx.doi.org/10.1016/j.apradiso.2014.12.001},
  DOI = {10.1016/j.apradiso.2014.12.001},
  journal = {Applied Radiation and Isotopes},
  publisher = {Elsevier BV},
  author = {Wooten,  A. Lake and Lewis,  Benjamin C. and Lapi,  Suzanne E.},
  year = {2015},
  month = feb,
  pages = {154–161}
}

@article{Graves2015Mnimmu,
  title = "{Novel Preparation Methods of 52Mn for ImmunoPET Imaging}",
  volume = {26},
  ISSN = {1520-4812},
  url = {http://dx.doi.org/10.1021/acs.bioconjchem.5b00414},
  DOI = {10.1021/acs.bioconjchem.5b00414},
  number = {10},
  journal = {Bioconjugate Chemistry},
  publisher = {American Chemical Society (ACS)},
  author = {Graves,  Stephen A. and Hernandez,  Reinier and Fonslet,  Jesper and England,  Christopher G. and Valdovinos,  Hector F. and Ellison,  Paul A. and Barnhart,  Todd E. and Elema,  Dennis R. and Theuer,  Charles P. and Cai,  Weibo and Nickles,  Robert J. and Severin,  Gregory W.},
  year = {2015},
  month = sep,
  pages = {2118–2124}
}

@article{Vanasschen2016Mn,
  title = "{Radiolabelling with isotopic mixtures of52g/55Mn(<scp>ii</scp>) as a straight route to stable manganese complexes for bimodal PET/MR imaging}",
  volume = {45},
  ISSN = {1477-9234},
  url = {http://dx.doi.org/10.1039/C5DT04270D},
  DOI = {10.1039/c5dt04270d},
  number = {4},
  journal = {Dalton Transactions},
  publisher = {Royal Society of Chemistry (RSC)},
  author = {Vanasschen,  Christian and Brandt,  Marie and Ermert,  Johannes and Coenen,  Heinz H.},
  year = {2016},
  pages = {1315–1321}
}

@article{Saar2018Mn,
  title = "{Anatomy,  Functionality,  and Neuronal Connectivity with Manganese Radiotracers for Positron Emission Tomography}",
  volume = {20},
  ISSN = {1860-2002},
  url = {http://dx.doi.org/10.1007/s11307-018-1162-6},
  DOI = {10.1007/s11307-018-1162-6},
  number = {4},
  journal = {Molecular Imaging and Biology},
  publisher = {Springer Science and Business Media LLC},
  author = {Saar,  Galit and Millo,  Corina M. and Szajek,  Lawrence P. and Bacon,  Jeff and Herscovitch,  Peter and Koretsky,  Alan P.},
  year = {2018},
  month = feb,
  pages = {562–574}
}

@article{Lewis2015Mn,
  title = "{52Mn Production for PET/MRI Tracking Of Human Stem Cells Expressing Divalent Metal Transporter 1 (DMT1)}",
  volume = {5},
  ISSN = {1838-7640},
  url = {http://dx.doi.org/10.7150/thno.10185},
  DOI = {10.7150/thno.10185},
  number = {3},
  journal = {Theranostics},
  publisher = {Ivyspring International Publisher},
  author = {Lewis,  Christina M. and Graves,  Stephen A. and Hernandez,  Reinier and Valdovinos,  Hector F. and Barnhart,  Todd E. and Cai,  Weibo and Meyerand,  Mary E. and Nickles,  Robert J. and Suzuki,  Masatoshi},
  year = {2015},
  pages = {227–239}
}

@article{Zhou2020Mn,
  title = "{Positron Emission Tomography–Magnetic Resonance Imaging Pharmacokinetics,  In Vivo Biodistribution,  and Whole-Body Elimination of Mn-PyC3A}",
  volume = {56},
  ISSN = {0020-9996},
  url = {http://dx.doi.org/10.1097/RLI.0000000000000736},
  DOI = {10.1097/rli.0000000000000736},
  number = {4},
  journal = {Investigative Radiology},
  publisher = {Ovid Technologies (Wolters Kluwer Health)},
  author = {Zhou,  Iris Yuwen and Ramsay,  Ian A. and Ay,  Ilknur and Pantazopoulos,  Pamela and Rotile,  Nicholas J. and Wong,  Alison and Caravan,  Peter and Gale,  Eric M.},
  year = {2020},
  month = nov,
  pages = {261–270}
}

@article{Hernandez2017Mndia,
  title = "{Radiomanganese PET Detects Changes in Functional $\beta$-Cell Mass in Mouse Models of Diabetes}",
  volume = {66},
  ISSN = {1939-327X},
  url = {http://dx.doi.org/10.2337/db16-1285},
  DOI = {10.2337/db16-1285},
  number = {8},
  journal = {Diabetes},
  publisher = {American Diabetes Association},
  author = {Hernandez,  Reinier and Graves,  Stephen A. and Gregg,  Trillian and VanDeusen,  Halena R. and Fenske,  Rachel J. and Wienkes,  Haley N. and England,  Christopher G. and Valdovinos,  Hector F. and Jeffery,  Justin J. and Barnhart,  Todd E. and Severin,  Gregory W. and Nickles,  Robert J. and Kimple,  Michelle E. and Merrins,  Matthew J. and Cai,  Weibo},
  year = {2017},
  month = may,
  pages = {2163–2174}
}

@article{Jensen2018Mn2,
  title = "{Remote-loading of liposomes with manganese-52 and in vivo evaluation of the stabilities of 52Mn-DOTA and 64Cu-DOTA using radiolabelled liposomes and PET imaging}",
  volume = {269},
  ISSN = {0168-3659},
  url = {http://dx.doi.org/10.1016/j.jconrel.2017.11.006},
  DOI = {10.1016/j.jconrel.2017.11.006},
  journal = {Journal of Controlled Release},
  publisher = {Elsevier BV},
  author = {Jensen,  Andreas I. and Severin,  Gregory W. and Hansen,  Anders E. and Fliedner,  Frederikke P. and Eliasen,  Rasmus and Parhamifar,  Ladan and Kjær,  Andreas and Andresen,  Thomas L. and Henriksen,  Jonas R.},
  year = {2018},
  month = jan,
  pages = {100–109}
}

@article{DeNardo2019Mn3,
  title = "{Radiation effective dose assessment of [51Mn]- and [52Mn]-chloride}",
  volume = {153},
  ISSN = {0969-8043},
  url = {http://dx.doi.org/10.1016/j.apradiso.2019.108805},
  DOI = {10.1016/j.apradiso.2019.108805},
  journal = {Applied Radiation and Isotopes},
  publisher = {Elsevier BV},
  author = {De Nardo,  Laura and Ferro-Flores,  Guillermina and Bolzati,  Cristina and Esposito,  Juan and Meléndez-Alafort,  Laura},
  year = {2019},
  month = nov,
  pages = {108805}
}

@article{Napieczynska2017Mn4,
  title = "{Imaging neuronal pathways with 52Mn PET: Toxicity evaluation in rats}",
  volume = {158},
  ISSN = {1053-8119},
  url = {http://dx.doi.org/10.1016/j.neuroimage.2017.06.058},
  DOI = {10.1016/j.neuroimage.2017.06.058},
  journal = {NeuroImage},
  publisher = {Elsevier BV},
  author = {Napieczynska,  Hanna and Severin,  Gregory W. and Fonslet,  Jesper and Wiehr,  Stefan and Menegakis,  Apostolos and Pichler,  Bernd J. and Calaminus,  Carsten},
  year = {2017},
  month = sep,
  pages = {112–125}
}

@article{CRMInte,
    author = {Ito, Kenji and Oka, Toshitaka and Kobayashi, Yoshinori and Shirai, Yasuharu and Wada, Ken’ichiro and Matsumoto, Masataka and Fujinami, Masanori and Hirade, Tetsuya and Honda, Yoshihide and Hosomi, Hiroyuki and Nagai, Yasuyoshi and Inoue, Koji and Saito, Haruo and Sakaki, Koji and Sato, Kiminori and Shimazu, Akira and Uedono, Akira},
    title = "{Interlaboratory comparison of positron annihilation lifetime measurements for synthetic fused silica and polycarbonate}",
    journal = {Journal of Applied Physics},
    volume = {104},
    number = {2},
    pages = {026102},
    year = {2008},
    month = {07},
    issn = {0021-8979},
    doi = {10.1063/1.2957074},
    url = {https://doi.org/10.1063/1.2957074},
}

@ARTICLE{KubatBAMS2024,
  title = "{Calibration of PALS System withCRM Materials for BiomedicalStudies}",
  volume = {20},
  ISSN = {1895-9091},
  url = {http://dx.doi.org/10.5604/01.3001.0054.9091},
  DOI = {10.5604/01.3001.0054.9091},
  number = {Special Issue},
  journal = {Bio-Algorithms and Med-Systems},
  publisher = {Index Copernicus},
  author = {Kubat,  Karol and Kapłon, {\L}ukasz and Moskal,  Paweł and Stępień,  Ewa},
  year = {2024},
  month = dec,
  pages = {35–39}
}

@ARTICLE{KorcylBAMS2014,
  title = "{Trigger-less and reconfigurable data acquisition system for positron emission tomography}",
  volume = {10},
  ISSN = {1895-9091},
  url = {http://dx.doi.org/10.1515/bams-2013-0115},
  DOI = {10.1515/bams-2013-0115},
  number = {1},
  journal = {Bio-Algorithms and Med-Systems},
  publisher = {Index Copernicus},
  author = {Korcyl,  Grzegorz and Moskal,  Paweł and Bednarski,  Tomasz and Białas,  Piotr and Czerwiński,  Eryk and Kapłon,  Łukasz and Kochanowski,  Andrzej and Kowal,  Jakub and Kowalski,  Paweł and Kozik,  Tomasz and Krzemień,  Wojciech and Molenda,  Marcin and Niedźwiecki,  Szymon and Pałka,  Marek and Pawlik,  Monika and Raczyński,  Lech and Rudy,  Zbigniew and Salabura,  Piotr and Gupta-Sharma,  Neha and Silarski,  Michał and Słomski,  Artur and Smyrski,  Jerzy and Strzelecki,  Adam and Wiślicki,  Wojciech and Zieliński,  Marcin and Zoń,  Natalia},
  year = {2014},
  month = mar,
  pages = {37–40}
}

@article{BAMSPalka2014,
  title = {A novel method based solely on field programmable gate array (FPGA) units enabling measurement of time and charge of analog signals in positron emission tomography (PET)},
  volume = {10},
  ISSN = {1895-9091},
  url = {http://dx.doi.org/10.1515/bams-2013-0104},
  DOI = {10.1515/bams-2013-0104},
  number = {1},
  journal = {Bio-Algorithms and Med-Systems},
  publisher = {Index Copernicus},
  author = {Pałka,  Marek and Moskal,  Paweł and Bednarski,  Tomasz and Białas,  Piotr and Czerwiński,  Eryk and Kapłon,  Łukasz and Kochanowski,  Andrzej and Korcyl,  Grzegorz and Kowal,  Jakub and Kowalski,  Paweł and Kozik,  Tomasz and Krzemień,  Wojciech and Molenda,  Marcin and Niedźwiecki,  Szymon and Pawlik,  Monika and Razyński,  Lech and Rudy,  Zbigniew and Salabura,  Piotr and Gupta-Sharma,  Neha and Silarski,  Michał and Słomski,  Artur and Smyrski,  Jerzy and Strzelecki,  Adam and Wiślicki,  Wojciech and Zieliński,  Marcin and Zoń,  Natalia},
  year = {2014},
  month = mar,
  pages = {41–45}
}

@article{SHIMAZOE2023,
  title = "{Improvement of double photon emission Compton imaging using angular and polarization correlations in cascade photons}",
  volume = {18},
  ISSN = {1748-0221},
  url = {http://dx.doi.org/10.1088/1748-0221/18/05/c05012},
  DOI = {10.1088/1748-0221/18/05/c05012},
  number = {05},
  journal = {Journal of Instrumentation},
  publisher = {IOP Publishing},
  author = {Kim,  Donghwan and Ueki,  Taisei and Rahman,  Agus Nur and Yan,  Linlin and Uenomachi,  Mizuki and Shimazoe,  Kenji and Takahashi,  Hiroyuki},
  year = {2023},
  month = may,
  pages = {C05012}
}

@article{Shimazoe2022,
  title = "{Imaging and sensing of pH and chemical state with nuclear-spin-correlated cascade gamma rays via radioactive tracer}",
  volume = {5},
  ISSN = {2399-3650},
  url = {http://dx.doi.org/10.1038/s42005-022-00801-w},
  DOI = {10.1038/s42005-022-00801-w},
  number = {1},
  journal = {Communications Physics},
  publisher = {Springer Science and Business Media LLC},
  author = {Shimazoe,  Kenji and Uenomachi,  Mizuki and Takahashi,  Hiroyuki},
  year = {2022},
  month = jan 
}

@article{ShimazoeBAMS2022,
  title = "{A double photon coincidence detection method for medical gamma-ray imaging}",
  volume = {18},
  ISSN = {1896-530X},
  url = {http://dx.doi.org/10.2478/bioal-2022-0080},
  DOI = {10.2478/bioal-2022-0080},
  number = {1},
  journal = {Bio-Algorithms and Med-Systems},
  publisher = {Index Copernicus},
  author = {Uenomachi,  Mizuki and Shimazoe,  Kenji and Takahashi,  Hiroyuki},
  year = {2022},
  month = dec,
  pages = {120–126}
}

@article{Li2024,
  title = "{Performance Evaluation of the uMI Panorama PET/CT System in Accordance with the National Electrical Manufacturers Association NU 2-2018 Standard}",
  volume = {65},
  ISSN = {2159-662X},
  url = {http://dx.doi.org/10.2967/jnumed.123.265929},
  DOI = {10.2967/jnumed.123.265929},
  number = {4},
  journal = {Journal of Nuclear Medicine},
  publisher = {Society of Nuclear Medicine},
  author = {Li,  Guiyu and Ma,  Wenhui and Li,  Xiang and Yang,  Weidong and Quan,  Zhiyong and Ma,  Taoqi and Wang,  Junling and Wang,  Yunya and Kang,  Fei and Wang,  Jing},
  year = {2024},
  month = feb,
  pages = {652–658}
}

@article{MercolliSc,
  title = "{First positronium lifetime imaging with scandium-44 on a long axial field-of-view PET/CT}",
  volume = {5},
  ISSN = {2673-8880},
  url = {http://dx.doi.org/10.3389/fnume.2025.1648621},
  DOI = {10.3389/fnume.2025.1648621},
  journal = {Frontiers in Nuclear Medicine},
  publisher = {Frontiers Media SA},
  author = {Mercolli,  Lorenzo and Steinberger,  William M. and Grundler,  Pascal V. and Moiseeva,  Anzhelika and Braccini,  Saverio and Conti,  Maurizio and Moskal,  Paweł and Rathod,  Narendra and Rominger,  Axel and Sari,  Hasan and Schibli,  Roger and Seifert,  Robert and Shi,  Kuangyu and Stepień,  Ewa Ł. and van der Meulen,  Nicholas P.},
  year = {2025},
  month = nov 
}

@article{ManishIEEE,
  title = "{First Positronium Imaging Using 44Sc With the J-PET Scanner: a Case Study on the NEMA-Image Quality Phantom}",
  ISSN = {2469-7303},
  url = {http://dx.doi.org/10.1109/TRPMS.2025.3621554},
  DOI = {10.1109/trpms.2025.3621554},
  journal = {IEEE Transactions on Radiation and Plasma Medical Sciences},
  publisher = {Institute of Electrical and Electronics Engineers (IEEE)},
  author = {Das,  Manish and Sharma,  Sushil and Beyene,  Ermias Yitayew and Bilewicz,  Aleksander and Choinski,  Jaroslaw and Chug,  Neha and Curceanu,  Catalina and Czerwinski,  Eryk and Eliyan,  Kavya Valsan and Hajduga,  Jakub and Jalali,  Sharareh and Kacprzak,  Krzysztof and Kaplanoglu,  Tevfik and Kaplon,  Lukasz and Kasperska,  Kamila and Khreptak,  Aleksander and Korcyl,  Grzegorz and Kozik,  Tomasz and Kubat,  Karol and Kumar,  Deepak and Venadan,  Anoop Kunimmal and Lisowski,  Edward and Lisowski,  Filip and Medrala-Sowa,  Justyna and Moyo,  Simbarashe and Mryka,  Wiktor and Niedźwiecki,  Szymon and Pandey,  Piyush and Parzych,  Szymon and Porcelli,  Alessio and Rachwal,  Bartlomiej and del Rio,  Elena Perez and R\"{a}dler,  Martin and Rominger,  Axel and Shi,  Kuangyu and Skurzok,  Magdalena and Stolarz,  Anna and Szumlak,  Tomasz and Tanty,  Pooja and Ardebili,  Keyvan Tayefi and Tiwari,  Satyam and Walczak,  Rafal and Stepien,  Ewa L. and Moskal,  Pawel},
  year = {2025},
  pages = {1–1}
}

@article{Shimazoe2bams2022,
  title = "{Multi-molecule imaging and inter-molecular imaging in nuclear medicine}",
  volume = {18},
  ISSN = {1896-530X},
  url = {http://dx.doi.org/10.2478/bioal-2022-0081},
  DOI = {10.2478/bioal-2022-0081},
  number = {1},
  journal = {Bio-Algorithms and Med-Systems},
  publisher = {Index Copernicus},
  author = {Shimazoe,  Kenji and Uenomachi,  Mizuki},
  year = {2022},
  month = dec,
  pages = {127–134}
}

@article{Nayak2009,
  title = "{Radioimmunoimaging with Longer-Lived Positron-Emitting Radionuclides: Potentials and Challenges}",
  volume = {20},
  ISSN = {1520-4812},
  url = {http://dx.doi.org/10.1021/bc800299f},
  DOI = {10.1021/bc800299f},
  number = {5},
  journal = {Bioconjugate Chemistry},
  publisher = {American Chemical Society (ACS)},
  author = {Nayak,  Tapan K. and Brechbiel,  Martin W.},
  year = {2009},
  month = jan,
  pages = {825–841}
}

@article{Likhithaswamy2022,
  title = "{Assessing the Quality of Long-Term Stored Tissues in Formalin and in Paraffin-Embedded Blocks for Histopathological Analysis}",
  volume = {10},
  ISSN = {2213-879X},
  url = {http://dx.doi.org/10.4103/JMAU.JMAU_53_20},
  DOI = {10.4103/jmau.jmau_53_20},
  number = {1},
  journal = {Journal of Microscopy and Ultrastructure},
  publisher = {Medknow},
  author = {Likhithaswamy,  Hariyabbe Rangaswamy and Madhushankari,  G. S. and Selvamani,  Manickam and Mohan Kumar,  K. P. and Kokila,  Ganganna and Mahalakshmi,  Saibaba},
  year = {2022},
  month = jan,
  pages = {23–29}
}

@article{Co_production,
  title = "{Cyclotron production and radiochemical separation of 55Co and 58mCo from 54Fe,  58Ni and 57Fe targets}",
  volume = {130},
  ISSN = {0969-8043},
  url = {http://dx.doi.org/10.1016/j.apradiso.2017.09.005},
  DOI = {10.1016/j.apradiso.2017.09.005},
  journal = {Applied Radiation and Isotopes},
  publisher = {Elsevier BV},
  author = {Valdovinos,  H.F. and Hernandez,  R. and Graves,  S. and Ellison,  P.A. and Barnhart,  T.E. and Theuer,  C.P. and Engle,  J.W. and Cai,  W. and Nickles,  R.J.},
  year = {2017},
  month = dec,
  pages = {90–101}
}

@article {Moyo2025for,
	author = {Moyo, Simbarashe and Wr{\'o}bel, Andrzej and Leszczy{\'n}ski, Bartosz and Skalska, Magdalena and St{\k e}pniewski, Jakub and Kope{\'c}, Grzegorz and Moskal, Pawe{\l} and St{\k e}pie{\'n}, Ewa},
	title = "{Fresh Paraformaldehyde Preserves Thrombus Biochemistry - Infrared Spectroscopy and Secondary Ion Mass Spectrometry Investigation of Pulmonary Artery Thrombi}",
	elocation-id = {2025.11.17.688840},
	year = {2025},
	doi = {10.1101/2025.11.17.688840},
	publisher = {Cold Spring Harbor Laboratory},
	URL = {https://www.biorxiv.org/content/early/2025/12/08/2025.11.17.688840},
	eprint = {https://www.biorxiv.org/content/early/2025/12/08/2025.11.17.688840.full.pdf},
	journal = {bioRxiv}
}

@misc{kubat2026exvivopositroniumimagingtissues,
      title="{First ex-vivo positronium imaging of tissues with modular J-PET scanner using $^{44}$Sc radionuclide}", 
      author={Karol Kubat and Manish Das and Sushil Sharma and Ermias Y. Beyene and Aleksander Bilewicz and Jarosław Choiński and Neha Chug and Catalina Curceanu and Eryk Czerwiński and Jakub Hajduga and Sharareh Jalali and Krzysztof Kacprzak and Tevfik Kaplanoglu and Łukasz Kapłon and Kamila Kasperska and Aleksander Khreptak and Grzegorz Korcyl and Tomasz Kozik and Deepak Kumar and Sumit Kumar Kundu and Anoop Kunimmal-Venadan and Bartosz Leszczyński and Edward Lisowski and Filip Lisowski and Justyna Mędrala-Sowa and Simbarashe Moyo and Wiktor Mryka and Szymon Niedźwiecki and Anand Pandey and Piyush Pandey and Szymon Parzych and Alessio Porcelli and Bartłomiej Rachwał and Martin Rädler and Magdalena Skurzok and Anna Stolarz and Tomasz Szumlak and Pooja Tanty and Keyvan Tayefi Ardebili and Satyam Tiwari and Kavya Valsan Eliyan and Rafał Walczak and Paweł Moskal and Ewa Ł. Stępień},
      year={2026},
      doi = {10.48550/arXiv.2602.07580},
      archivePrefix={arXiv},
      primaryClass={physics.med-ph},
      url={https://arxiv.org/abs/2602.07580}, 
}

\section*{Acknowledgements}

We acknowledge the technical support of Andrzej Heczko, Marcin Kajetanowicz, Dr. Piotr  Kapusta, Wojciech Migdał, and Adam Mucha. We would like to thank Dr. Grzegorz Grudzień for providing access to myxoma and adipose samples. We acknowledge the support of Claire Deville and Kristina Søborg Pedersen from the Technical University of Denmark. We acknowledge support from the National Science Centre of Poland through grants MAESTRO no. 2021/42/A/ST2/00423 (P.M.), OPUS no. 2021/43/B/ST2/02150 (P.M.), OPUS24+LAP no. 2022/47/I/NZ7/03112 (E.Ł.S.) and SONATA no. 2023/50/E/ST2/00574 (S.S.), the Ministry of Science and Higher Education through grant no. IAL/SP/596235/2023 (P.M.), the SciMat and qLife Priority Research Areas budget under the program Excellence Initiative – Research University at Jagiellonian University (P.M. and E.Ł.S.), the Research Support Module as part of the Excellence Initiative – Research University program at Jagiellonian University (M.D.), European Union within the Horizon Europe Framework Programme (ERC Advanced Grant POSITRONIUM no. 101199807) and PRISMAP via Project\_1729020993\_aniX3. We also acknowledge Polish high-performance computing infrastructure PLGrid (HPC Center: ACK Cyfronet AGH) for providing computer facilities and support within computational grant no. PLG/2024/017688 and PLG/2025/018762.

\section*{Author contributions statement}

The J-PET scanner, the techniques of the experiment, and this study were conceived by P.M. The data analysis was conducted by M.D. Signal selection criteria were developed by M.D, P.M. and S.S., applied by M.D., and verified by S.S. The samples were prepared by K. Kubat. J.C., N.Razzaq, R.W. and A.B. designed and carried out the irradiation and processing of the $^{54}$Fe targets. Authors M.D., S.S., E.Y.B., A.B., J.C., N.C., C.C., E.C., J.H., S.J., K.Kacprzak, T.K., Ł.K., K. Kasperska, A.K., G.K., T.K., K. Kubat, D.K., S.K.K, A.K.V.,
E.L., F.L., J.M., S.M., W.M., S.N., A.P., P.P., S.P., A.P., B.R., M.R., N. Rathod, N. Razzaq, A.R., K.S., M. Skurzok, M. Słotwiński, A.S., T. S., P.T., K.T.A., S.T., K. V. E.,
R.W., E.Ł.S, and P.M. participated in the construction, commissioning, and operation of the experimental setup, as well as in the data-taking campaign and data interpretation. K.Kacprzak took part in developing the J-PET analysis and simulation framework. M.Skurzok and K.Kacprzak performed the timing calibration of the detector. E.C. developed and operated short- and long-term data archiving systems and the computer center of J-PET. S.S. established the relation between energy loss and TOT and the dependence of detection efficiency on energy deposition. P.M. and E.Ł.S conceptualized the study, secured the main financing and supervised the whole project. The results were interpreted by P.M., E.Ł.S, S.S., and M.D. The manuscript was prepared by P.M., E.Ł.S, S.S., and M.D. and was then edited and approved by all authors.

\section*{Competing interests}

The authors declare the following commercial interests/personal relationships which may be considered as potential competing interests with the work reported in this paper: Paweł Moskal is an inventor on a patent related to this work. [Patent nos.: (Poland) PL 227658, (Europe) EP 3039453, and (United States) US 9,851,456], filed (Poland) 30 August 2013, (Europe) 29 August 2014, and (United States) 29 August 2014; published (Poland) 23 January 2018, (Europe) 29 April 2020, and (United States) 26 December 2017.

Other authors declare that they have no known conflicts of interest in terms of competing commercial interests or personal relationships that could have an influence or are relevant to the work reported in this paper.

\end{document}